\documentclass[useAMS,usenatbib]{mnras}
\usepackage{amsmath}
\usepackage{ae,aecompl}
\usepackage{enumitem}
\usepackage{graphicx}
\usepackage{amssymb}
\usepackage{color}
\usepackage{threeparttable}
\usepackage{tabularx}
\usepackage{changes}           

\usepackage{float}
\usepackage{fancyref}
\usepackage{comment}
\usepackage[T1]{fontenc}
\usepackage{mathptmx}
\newcommand{\msun}{\mbox{${\rm M_\odot}$}}
\newcommand{\myr}{\mbox{${\rm M_\odot~yr^{-1}}$}}
\newcommand{\mign}{\mbox{$M_{\rm ign}$}}
\newcommand{\mej}{\mbox{$M_{\rm ej}$}}

\newcommand{\isot}[2]{${}^{#1}$#2}

\newcommand{\rsun}{\mbox{$\mathrm{R_{\odot}}$}}
\newcommand{\lsun}{\mbox{$\mathrm{L_{\odot}}$}}
\newcommand{\mdot}{\mbox{$\dot{\mathrm{M}}$}}
\newcommand{\dotm}{\mbox{$\dot{\mathrm{M}}$}}
\newcommand{\mdota}{\mbox{$\dot{\mathrm{M}}_{\mathrm{accr}}$}} 
 
\newcommand{\pyr}{\mbox{{\rm yr$^{-1}$}}}
\newcommand{\mwd}{\mbox{{\rm M$_{\rm WD}$}}}
\newcommand{\ledd}{\mbox{{\rm L$_{\rm Edd}$}}}
\def\apgt{{\raise-.5ex\hbox{$\buildrel>\over\sim\,$}}}
\def\aplt{{\raise-.5ex\hbox{$\buildrel<\over\sim\,$}}}

\title[]{Comprehensive models of novae at metallicity $Z = 0.02$ and $Z = 10^{-4}$ }

\author[H.-L. Chen et al.]{Hai-Liang Chen,$^{1,2,3,4}$\thanks{E-mail:chenhl@ynao.ac.cn}
T. E. Woods,$^{5}$
L. R. Yungelson,$^{6}$
Luciano Piersanti,$^{8,9}$
\newauthor
M. Gilfanov,$^{2,7,10}$ 
and Zhanwen Han$^{1,3,4}$
\\
$^{1}$Yunnan Observatories, Chinese Academy of Sciences, Kunming, 650011,China\\
$^{2}$Max Planck Institute for Astrophysics, Karl-Schwarzschild-Str. 1, Garching b. M{\" u}nchen 85741, Germany\\
$^{3}$Key Laboratory for the Structure and Evolution of Celestial Objects, Chinese Academy of Sciences, Kunming 650011, China\\
$^{4}$Center for Astronomical Mega-Science, Chinese Academy of Science, Beijing 100012, China\\
$^{5}$Institute of Gravitational Wave Astronomy and School of Physics and Astronomy, University of Birmingham, Edgbaston, Birmingham B15 2TT, United Kingdom\\
$^{6}$Institute of Astronomy, RAS, 48 Pyatnitskaya Str., 119017 Moscow, Russia\\
$^{7}$Space Research Institute of Russian Academy of Sciences, Profsoyuznaya 84/32, 117997 Moscow, Russia\\
$^{8}$INAF -- Osservatorio Astronomico d'Abruzzo, via Mentore Maggini, snc, I-64100, Teramo, Italy\\
$^{9}$INFN -- Sezione di Perugia, via Pascoli, Perugia, Italy \\
$^{10}$ Kazan Federal University, Kremlevskaya str. 18, 420008 Kazan, Russia
}

\date{Accepted XXX. Received YYY; in original form ZZZ}

\begin{document}
\label{firstpage}

\pagerange{\pageref{firstpage}--\pageref{lastpage}}
\maketitle

\begin{abstract}
Novae are the observational manifestations of thermonuclear runaways on the surface of accreting white dwarfs 
(WDs). Although novae are an ubiquitous phenomenon,
their properties at low metallicity 
are not well understood. 
Using the publicly-available stellar evolution code Modules for Experiments in Stellar Astrophysics (\textsc{mesa}), we 
model the evolution of accreting carbon-oxygen WDs and consider models
which accrete matter with metallicity Z=0.02 or $10^{-4}$. We  consider both models 
without mixing and with matter enriched by 
 CO-elements assuming that mixing occurs 
in the process of accretion 
(with mixing fraction 0.25).
We present and contrast ignition mass, ejected mass, recurrence period and maximum luminosity of novae for 
different WD masses and accretion rates for these metallicities and mixing cases. 
We find that models with Z = 0.02 have ignition masses and recurrence
periods smaller than models with low Z, while the ejected mass and maximum luminosity are larger. 
Retention efficiency during novae outbursts decreases with increasing metallicity.
In our implementation, inclusion of mixing at the H/He interface reduces 
accreted mass, ejected mass and recurrence period as compared to
the no-mixing case, while the maximum luminosity becomes larger. 
Retention efficiency is significantly reduced, becoming negative in most of our models.
For ease of use, we provide a tabular summary of our results.
\end{abstract}


\section{Introduction}
\label{sec:int}

It is well established that novae occur in binaries hosting accreting white dwarfs, if the accretion rate is lower than a critical threshold \citep[see, e.g.,][and references therein]{warn03}.  Accreted matter is compressed at the bottom of the envelope, leading to an 
increase in temperature.  Eventually 
nuclear hydrogen burning is ignited in the degenerate matter, leading to a thermonuclear runaway (TNR). This is
observable as a nova if the outburst results in mass loss. 

On evolutionary timescales, all nova outbursts recur \citep{pz78}.
If the accretion rate is larger than a critical value, there is a regime in which
hydrogen may burn stably in the envelope 
of the WD. If, however, the accretion rate is larger than the maximum stable
burning rate, then no hydrostatic solutions are possible for a thin burning shell (see discussion in~\citealt{sb07}). In this case, the ensuing evolution remains uncertain. A number of studies \citep[e.g.,][]{nns79,1988ApJ...324..355I,cit98} have found that, at such high accretion rates and nuclear-burning luminosities, the WD envelope will expand to red giant dimensions.
\citet{hkn96} argued, however, that a peak in the opacity due to iron will lead to an 
 optically thick wind, 
moderating the accretion rate and preventing significant expansion of the WD envelope.
 
Novae have been observed in the disk and bulge of the Galaxy and in external 
galaxies of different morphological types 
\citep[see, e.g., ][]{scpb+14,2017ApJ...834..196S}, 
as well as in Galactic and extragalactic globular clusters  
\citep[see, e.g.,][ and references therein]{sd95,khh13,cspn+15}.
Novae in different populations exhibit different characteristics. This may be
interpreted as a
consequence of a dependence of the masses of WDs and                  
their companions on the metallicity \citep[e.g., ][]{unyw99,dgsl+15} and typical age of         
stellar populations
\citep[e.g.,][]{dell02,scpb+14}. 

A link between the properties of classical Novae and metallicity has long been established.
\citet{sts78} found that substantial enrichment by carbon is necessary in order
to release sufficient energy for the rapid ejection of matter, while enrichment of accreted envelopes 
by $\beta^+$ unstable isotopes, which decay and release energy, facilitates mass ejection \citep[see also ][for the latest review]{2016PASP..128e1001S}.
Through the opacity, metallicity also influences the amount of energy 
retained in the burning layer, and hence the rate of temperature growth and the 
amount of ejected mass. As well, the light-curves of Novae have been shown to become slower with decreasing metallicity \citep{kato97,khh13}. Observationally, it was discovered that "all
novae for which reasonable abundance data are available appear
to be enriched in either helium or heavy elements, or both", while  
the presence of Ne, Na, Mg, and Al in ejecta suggested the existence of ONe accretors \citep{tl86}. 
From these observations and model data it was inferred that 
matter in the burning layers of novae is enriched by the matter 
from the cores of WDs. 

Several 2D and 3D simulations of TNR
\citep[e.g.][]{gl95,
cjgc+10,cjgs16,cjs18,glt12,2014ASPC..490..275J,2017gacv.workE..66S}
have demonstrated that Kelvin-Helmholtz instabilities arising when a TNR has already fully developed can 
lead to enrichment of the 
envelope to levels consistent with observations. 
Motivated by these studies, \citet{dhbp13} applied a convective boundary mixing 
algorithm implemented in the 1-D stellar evolution code \textsc{mesa}, and succeeded in
reproducing an enhancement of metal abundances in the 
ejected matter up to $Z = 0.29$ for a $1.20\;\msun$ CO WD, commensurate with observed values. 2D simulations by \citet{glt12} and \citet{cjgs16,cjs18}
have shown that the mixing process may  
depend on the WD composition. In particular, in ONeMg WDs the scale of mixing is larger than in CO WD,
leading to more energetic outbursts. It should be noted that these 2D and 3D 
simulations 
have only been carried out for a quite limited number of combinations of WD masses and accretion rates. 

In hydrodynamic studies \citep{1987Ap&SS.131..431P}, it was found that mass loss over the course of a full Nova eruption cycle is driven by dynamical acceleration of the matter leading to shock-ejection, followed by phases of continuous mass loss via optically thick winds and nebular mass-loss. 
Mass loss terminates when the mass of the
envelope declines below a critical limit and nuclear burning is extinguished 
\citep{fuji82}. 
In hydrostatic computations similar to the present study,
approximations to this mechanism are applied, such as mass loss by opacity-peak-driven optically thick sub-Eddington 
winds \citep[e.g., ][]{kato97}, optically thick radiatively driven
super-Eddington winds (SEW)
\citep{2002AIPC..637..259S}, mass-loss via common envelope \citep[e.g.,][]{lsbt90} or
loss of all expanding WD matter that 
crosses the Roche lobe radius \citep[e.g.,][]{wbbp13}.
Ultimately, understanding which of these processes (or what combinations) are principally responsible for driving mass loss must be informed by future observations.

It was speculated already in early studies 
\citep[e.g., ][]
{scha63,tc71,1982ApJ...260..249F,1984ApJ...283..241M,1986ApJ...310..222P,sss88} 
that WDs experiencing nova outbursts may retain a fraction of accreted matter and ultimately grow to become the
progenitors of type Ia supernovae (SNe~Ia) upon reaching the Chandrasekhar mass limit. 
 The fraction of accreted matter which is not ejected by novae is referred to as the retention efficiency, defined as
\begin{equation}
\eta = \frac{M_{\rm acc}^{tot}-M_{\rm ej}}{M_{\rm acc}^{tot}},
\label{eq:reten}
\end{equation}
where $M_{\rm acc}^{tot}$ is the  mass accreted by a WD over a single nova cycle, and $M_{\rm ej}$ is the ejected mass.
Note that $\eta$ may be negative, if the WD is eroded as a result of an outburst.

Recent theoretical and observational efforts have placed strong limits on the amount of matter which may be accumulated by accreting WDs in the steady-burning regime, particularly in the context of SN Ia progenitors \citep[e.g., ][]{gb10,wg14,jwgs+16,dhbr+17,wgbg17,wgbg18,GW19,KGSWV19}. The contribution of accreting WDs in the nova regime has also been strongly constrained for metallicities typical of the local Universe \citep[see e.g.,][]{SG2015}, however relatively little is known for the low metallicity case.
In particular, improving our understanding of 
the mass retention efficiency remains a critical issue 
\citep[see, e.g.][]{btn13,mmn14,py14}. 

The retention efficiency may 
be expected to be higher in low-Z environments, since the opacity of stellar matter
(which sets the pace of mass-loss) is then lower, and the effects of expansion due to radiation pressure are smaller.  \citet{pcit00} have investigated the evolution of low-mass accreting WDs 
(\mwd$\leq 0.68\,\mathrm{M_{\odot}}$) 
with  metallicities $Z = 0.02, 10^{-3}, 10^{-4}$. They found that at low metallicity
hydrogen may burn stably at lower accretion rates than at solar metallicity,
and the
mass necessary for a
TNR at a given \mdot\ is larger than for solar metallicity.  
The latter result was confirmed by subsequent studies \citep{ssts00,jghg07,sb07} who studied accretion onto 
more massive WDs for a range of sub-solar metallicities.  

The aim of this paper is to systematically study the characteristics of novae 
at approximately solar 
and very low metallicity, and thereby draw
conclusions on  
the influence of metallicity on nova properties.  
We also investigate the significance of enrichment of the accreted H-rich layer by the matter from the underlying CO-core (``mixing''). 
In this work, we are primarily interested 
in the conditions for steady burning of hydrogen at the surface of accreting WDs, the amount of mass to be accreted to trigger an outburst,
 and in the amount of mass eventually ejected. This is motivated by our eventual goal
of 
modelling of nova populations in different environments, and determining the precursors of SNe~Ia.

For this work, we calculated a grid of 
accreting CO WD models for a range of masses 
\mwd$= 0.51 - 1.30\;\mathrm{M_{\odot}}$, 
accretion rates $\dot{M} = 10^{-10} - 10^{-6}\;M_{\odot}\;{\rm yr}^{-1}$, and 
metallicities $Z = 0.02, 10^{-4}$. These two values of Z may be considered representative of metal-rich and metal-poor environments.
For both values of Z, we consider models with a simplified mixing approximation and without it.  

The paper is structured as follows. In Section~\ref{sec:sim}, we describe the approach we used to obtain initial WD models, and our assumptions 
in modelling the evolution of accreting WDs. In Section~\ref{sec:res}, we present 
our results for the ``steady burning strip'', 
describe in detail a full nova cycle 
at low metallicity and its differences with a cycle for high Z, and present physical characteristics of Novae at  
metallicities $Z = 0.02$ and $Z = 10^{-4}$. 
Uncertainties and implications of our results are addressed in 
Section~\ref{sec:disc}. 
Finally, we summarise our results and conclude in Section~\ref{sec:sum}.

\section{Simulation Assumptions}
\label{sec:sim}

\subsection{Initial WD models}

In our study, we use the stellar evolution code \emph{Modules for Experiments in Stellar Evolution} (\textsc{mesa} version 4906; \citealt{pbdh+11,pcab+13,pmsb+15})\footnote{\textsc{mesa} web page: \texttt{http://mesa.sourceforge.net}.}. 
To obtain the initial WD models, we evolve stars with different initial masses and different metallicities, from the Zero-Age Main-Sequence to the Asymptotic Giant Branch, halting at the stage when the stellar CO-core mass is 
close to the desired WD mass.
At that point, 
we impose a very high mass loss rate to remove the envelopes of the stars. This is the canonical approach
adopted
by many previous studies \citep{dhbp13,mccd+13,wbbp13,wang2018}.         

In this way, we produce CO WDs models with mass M~=~$0.51, 0.60, 0.70, 0.80,0.90, 1.00\;{\rm M_{\odot}}$.
For CO WDs with mass $M > 1.00\;{\rm M_{\odot}}$, we make WD models by allowing a $1.0\;{\rm M_{\odot}}$ WD to accrete C/O matter
until its total mass reaches $1.10\;{\rm M_{\odot}}$, $1.20\;{\rm M_{\odot}}$, or $1.30\;{\rm M_{\odot}}$. In this work, we do not explore ONe WD models.
This is because, after reconsideration of observational data, \citet{lt94} found that the fraction of novae with ONe WDs may be 
much lower than previously suggested. More recently, \citet{cwyg+16} obtained a similar conclusion
from their population synthesis study. 

After each initial WD model was produced, it was allowed to cool to a lower temperature. \citet{tb04} found that the equilibrium core temperatures 
of WDs in typical cataclysmic variables are below $10^{7}$\;K.
In addition, \citet{cwyg+16} found that the low temperature models from \citet{ypsk05} are preferred 
when compared to observational data of novae in the M31 galaxy.             
 Here we neglect the possible effect of H/He shells on the cooling of WDs and their central temperatures.
We assume the WD central temperature to be $10^{7}$\;K and cool every initial white dwarf model until 
its central temperature becomes very close to this value. 
The initial WD models have a thin He and H shell on their surface.
 The H shell masses are between $10^{-6}-10^{-4}\;M_{\odot}$ and the He shell masses are between $10^{-4}-10^{-2}\;M_{\odot}$.

\subsection{Nova outburst calculation}
\label{subsec:nova_cal}

To simulate nova outbursts, we follow the evolution of models using \textsc{mesa} for different combinations of initial WD mass and fixed accretion rate. We have computed nova models with metallicities $Z = 0.02$ and $10^{-4}$ and  
mixing fractions 0.0 (``no mixing case'') and 0.25 (``mixing case'').
In order to better understand the dependence of nova properties on metallicity, we have also computed several models with 
metallicities ranging from $Z = 10^{-5}$ to $Z = 0.08$
(see Figs.~\ref{fig:nova_mul_Z},~\ref{fig:eff_com_mul_z} below).
 For any given metallicity $Z$, the hydrogen mass fraction is computed as 
 $X = 0.76-3.0Z$, $Y=0.24+2.0Z$;  note that $\Delta Y/\Delta Z$=2 is 
appropriate for abundances from primordial to solar \citep{psht+98}.
Initial relative abundances of different isotopes in the  matter were implemented 
following \citet{lodd03}. 
In our study, we model mass accretion rates ranging from $10^{-10}\;{\rm M_{\odot}\; {yr}^{-1}}$ to $10^{-6}\;{\rm M_{\odot}\;{yr}^{-1}}$.
We did not consider accretion rates lower than $10^{-10}\;{\rm M_{\odot}\;{yr}^{-1}}$, as such accretion rates correspond to CVs that have evolved beyond the period minimum (they have ``bounced''). Though several novae with orbital periods 
between about 2\,hr and 80\, min are known \citep{2011yCat....102018R}, it is hard to
identify their evolutionary stage. While extant models predict the possibility of novae
among bouncers \citep{ypsk05}, they are certainly rare, due to their long recurrence periods, and therefore hardly contribute to the statistics of observed novae  \citep{cwyg+16}. 

In our calculations, we do not consider the rotation of WDs and convective overshooting.
Given that mixing processes cannot be investigated self-consistently 
in 1D stellar models, we do not 
include mixing with the core via a physical mechanism. Instead, following 
\citet{pstw+95}, we assume that accreted matter is enriched in C, O, and Ne. 
The enrichment (mixing fraction) may be constrained by observations of the abundances of heavy elements in the ejecta of novae.
However, it is well known that estimates of the
metallicities of the latter 
might  be rather uncertain  
(e.g. \citet{gtws98,kidj+13}; see also Jose and Shore in \citet{be08}.)
Even for the same nova, abundances derived from different observations can vary by 
a factor of several.
Motivated by this circumstance, \citet{kidj+13} proposed using the line ratios 
$\Sigma{\rm CNO/H}$, Ne/H, Mg/H, Al/H and Si/H as measures of the 
mixing fraction, and found that the latter should be $\le 0.25$ (by mass).
Note, however, that \citet{kidj+13} focused on ONe WDs only and the latter  have greater mixing compared to CO WDs \citep{cjs18}. Therefore, 
a mixing fraction of 0.25 can be taken as an upper limit for CO WDs. 
We use this value of the mixing fraction in our calculations in order to test the maximum effect that mixing may have upon nova explosions.
It is worth noting, however, that the mixing fraction should depend on the WD mass and accretion rate. 

In models with mixing, the accreted matter is 
enriched in CNO cycle catalysts. In principle, 
the abundances of matter in the underlying white dwarfs should depend on their progenitor mass, the white dwarf mass itself, overshooting, and mass loss
during the preceding evolution. Here, however, we make the simplifying assumption of
neglecting the differing chemical abundances among WDs of differing masses, and  
assume that the chemical composition of all underlying white dwarfs 
is $X(^{12}{\rm C}) = 0.41$, $X(^{16}{\rm O}) = 0.57$, $X(^{22}{\rm Ne}) = 0.02$ 
for $Z = 0.02$ models and $X(^{12}{\rm C}) = 0.6354$, $X(^{16}{\rm O}) = 0.3645$, 
$X(^{22}{\rm Ne}) = 10^{-4}$ for $Z = 10^{-4}$ models.
As a sanity check for this assumption, we computed toy models with $X(^{12}{\rm C}) = 0.88$, $X(^{16}{\rm O}) = 0.10$, 
$X(^{22}{\rm Ne}) = 0.02$ and $X(^{12}{\rm C}) = 0.10$, $X(^{16}{\rm O}) = 0.88$, $X(^{22}{\rm Ne}) 
= 0.02$. Our results (together with those for our ``standard'' model) are presented in 
Table~\ref{tab:enr}.  They agree with the inference of \citet{hjci96} that the abundance of \isot{12}{C} is
crucial for the development of a TNR, which develops more 
rapidly and has lower \mign\ with increasing \isot{12}{C} in the accreted matter
\citep[see also, e.g., ][]{2017gacv.workE..66S}.   
However, the difference in \mign\ and \mej\ between the 
extreme cases does not exceed about 30\% and is $\leq 16$\% for $L_{max}$. 
This difference cannot be considered
significant given the present uncertainties in modeling novae.

\begin{table}
\caption{Comparison of models with different enrichment of accreted matter. Masses are in \msun.}
\begin{tabular}{p{7mm}ccccc}\hline
\mwd & X(\isot{12}{C}) & X(\isot{16}{O}) & \mign  & \mej & $L_{\rm max}$/\lsun \\
\hline
0.51 & 0.41   & 0.57 & $1.05 \cdot10^{-4}$  & $9.67 \cdot10^{-5}$    & $4.22 \cdot10^4$ \\
0.51 & 0.88   & 0.10 & $8.67 \cdot10^{-5}$  & $7.85\cdot 10^{-5}$    & $4.09\cdot 10^4$ \\
0.51 & 0.10   & 0.88 & $1.15\cdot 10^{-4}$  & $1.03 \cdot10^{-4}$    & $4.07 \cdot10^5$   \\
1.0  & 0.41   & 0.57 & $1.25\cdot 10^{-5}$  & $1.19\cdot 10^{-5}$    & $1.24 \cdot10^4$ \\
1.0  & 0.88   & 0.10 & $1.16\cdot 10^{-5}$  & $1.15\cdot 10^{-5}$    & $1.16\cdot 10^4$   \\
1.0  & 0.10   & 0.88 & $1.34\cdot 10^{-5}$  & $1.30 \cdot10^{-5}$    & $1.34\cdot 10^4$ \\
\hline
\end{tabular}
\label{tab:enr}
\end{table}

We use the MESA nuclear network \texttt{cno\_extras\_o18\_to\_mg26\_plus\_fe56}, which includes 29 isotopes, from H to \isot{26}{Mg} and \isot{56}{Fe}, linked by 75 nuclear
processes. This network is based on the one derived by \citet{wbbp13} and includes $pp$ and $pep$ 
chains, as well as cold and hot CNO and NeNa cycles. It represents the minimum network needed to accurately evaluate the energy delivered by nuclear burning during a 
full nova cycle 
and, hence, it allows the correct determination of the physical evolution of an accreting CO WD.\footnote{
We note that this network does not include the MgAl cycle, active for temperatures above $\sim 2.5\times 10^8$ K; this cycle is important for deriving the
nucleosynthesic yield, but provides a negligible contribution to the nuclear energy production.}

In \textsc{mesa},  an adaptive grid method for spatial discretization is used.
 As a compromise between numerical accuracy and computational time, in the inner part of the model where Lagrangian coordinates are used, we determine the number of grid points using the criterion  
\texttt{mesh\_delta\_coeff = 0.25}. This increases the number of mesh-points by a factor close to 4, compared to
the number set by default; in the outer part of the model, where the cell size is defined as a fraction of the
total mass (dq), we increased the number of grid points by setting \texttt{max\_surface\_cell\_dq = 1d-12}.                                                                                       
With these values, the typical number of zones in our model is around $6\times10^{3}-10^{4}$.
In order to verify that our choice of mass grid does not introduce an artificial mixing at the boundaries between regions with different chemical composition, 
thus producing inaccurate results, 
we performed test calculations for  \mwd=1\,\msun and
\mdot=$10^{-8}$\,\msun/yr
 with \texttt{mesh\_delta\_coeff = 1.0}, \texttt{max\_surface\_cell\_dq = 1d-12} and
\texttt{mesh\_delta\_coeff = 0.25}, \texttt{max\_surface\_cell\_dq = 1d-10}. Variations of accreted mass prior to outburst and ejected mass
did not exceed 12\%, while the recurrence time between the outbursts changed by less than 8\% and $\log{L_{max}}$ varied by only 0.01.

Mass loss rates during outbursts are computed using the default prescription given 
in \textsc{mesa},  super-Eddington wind (SEW): 
\begin{equation}
    \dot{M} = -2\frac{L-L_{\rm edd}}{v^{2}_{\rm esc}},  \qquad \quad\\
    L_{\rm edd} = \frac{4\pi GMc}{\overline{\kappa}},
\label{eq:wind_rate}
\end{equation}
 where $v_{\rm esc}$ is the escape velocity  at the photosphere of the accreting WD, $L_{\rm edd}$ is the Eddington luminosity 
and $\overline{\kappa}$ is the mass-weighted-mean Rosseland opacity of the outer layers of the WD  
(optical depth $\leq$ 100).\footnote{ A test run for \mwd=1.2\,\msun, $\mdot =10^{-8}\,{\rm M_\odot yr^{-1}}$ has shown that variation of this optical depth from $\tau=100$ to $\tau = 30$ resulted in an increase of 
the accreted mass prior to ignition by about 8\% 
and the ejected mass by only $\simeq$11\%, i.e., the dependence on the actual optical depth over which the
opacity is averaged does not have a substantial effect.}  
Equation~(\ref{eq:wind_rate}) is inspired by the finding 
that energy injection slightly below the point where the local escape velocity
exceeds the sonic speed may drive super-Eddington winds 
\citep{1985ApJ...289..634Q,1986ApJ...302..519P,2002AIPC..637..259S}; 
see also, e.g., 
\citet{2016MNRAS.458.1214Q} for a more recent discussion.  We will discuss the influence of the mass loss prescription on our results in greater detail in \ref{subsec:inf_ml}. 

In an effort to assess the steady-state behaviour of novae (i.e., after many outbursts), in our \textsc{mesa} calculations we follow the evolution of each accreting WD through $\simeq\,1-100$ novae outbursts, depending principally on the convergence  of each accreting white dwarf model.                                                                                        


\section{Results}
\label{sec:res}

\subsection{Steady burning regime}

\begin{figure}   
    \includegraphics[width=\columnwidth]{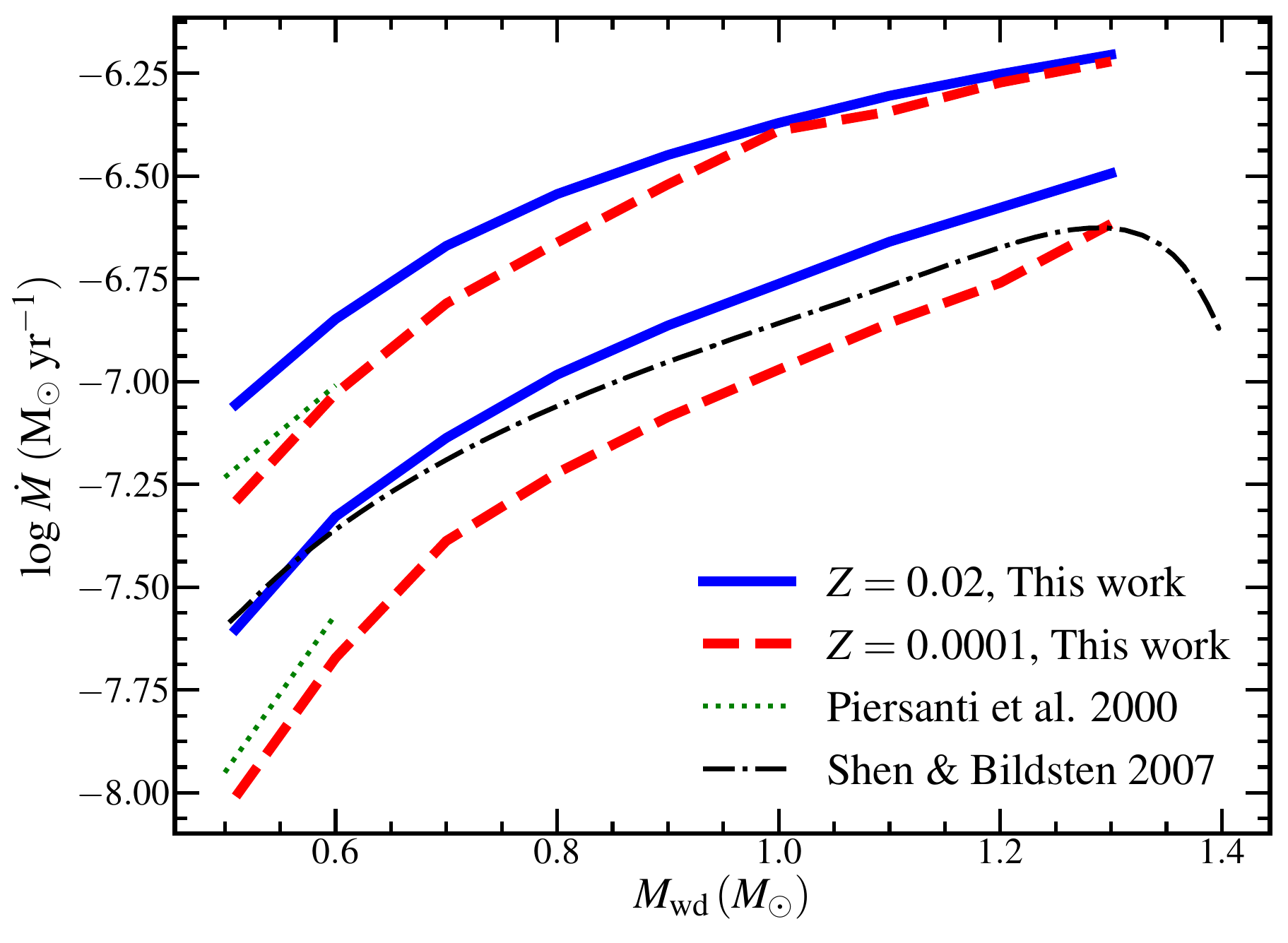}
    \caption{Steady burning regimes for $Z = 0.02$ (blue thick solid line) and $Z = 10^{-4}$ (red thick lines) 
     for our models without mixing in the \mwd-$\dot{M}$ plane. The thin dotted line represents       
     the steady burning regime for $Z = 10^{-4}$ from \citet{pcit00} and the thin dash-dotted line 
     represents the lower boundary of the steady burning regime for $Z = 10^{-4}$ from \citet{sb07}.
     The data and fitting formulae of these boundaries from our calculation are presented in Appendix A.
     }
    \label{fig:snbb_diff_Z}
\end{figure}

In Fig.~\ref{fig:snbb_diff_Z}, we show  the limits of the steady-state burning 
regimes for models of accreting WD  
with  $Z = 0.02$ and $Z = 10^{-4}$ and without mixing.
The data on these boundaries in our calculations are presented in Tables~\ref{tab:snbb_z2m2} and~\ref{tab:snbb_z1m4}. 
Fitting formulae for the boundaries are also provided in Appendix~A. In the steady-state H-burning 
regime, the bolometric luminosity of a WD remains constant.  Accreting WDs with accretion rates larger than the maximum
steady-burning rate greatly expand, while 
in the steady-burning regime their radii do not increase significantly.
Below the lower boundary, H-burning is unstable and the bolometric luminosity experiences quasi-regular oscillations (\citealt{pz78,iben82}, see also Fig.1 in \citealt{mccd+13}).
As is well known, there is no gradual transition between
stable and unstable burning, i.e., flashes appear below the limiting accretion rate with non-zero amplitude.
The plot shows that the steady burning boundary is lower at lower metallicity. 
The reason for this is straightforward:  the limits for steady burning  are determined by the requirement that H-burning energy balances the radiative losses from the surface along the high 
luminosity branch in the HRD-loop
(see e.g.~\citealt{pcit00}). As the luminosity level of this branch depends mainly on the mass 
of the CO core underlying the burning shell, 
it is about the same for all metallicities. For lower metallicity, however, the CNO abundance in the models without mixing is lower,
leading to a lower H-burning rate. On the other hand, the condition for the steady burning regime is that the accretion rate is equal 
to the H-burning rate.  This explains why the limits of the steady burning regime are located at lower values of $\dot{M}$ for lower $Z$.

In Fig.~\ref{fig:snbb_diff_Z}, we also compare our results for the steady burning regimes at 
$Z = 10^{-4}$ with previous studies. 
Note that \citet{pcit00}  considered only low mass WDs and \citet{sb07} present only the lower boundary of the steady burning 
regime for $Z = 10^{-4}$. Our results are consistent with the results of \citet{pcit00}. The lower boundary of the steady burning 
regime in this work, however, is below that obtained by \citet{sb07}. 
This is mainly due to the fact that \citet{sb07} adopted a one-zone approximation, in which the properties of the H-shell are defined at the bottom of the shell,
while the maximum of energy generation  occurs at larger Lagrangian coordinate.
Given that a part of the
nuclear burning energy can be absorbed by the accreted H-layer itself and
massive WDs have thinner H-layers, less energy will be absorbed. This is why the discrepancy for massive WDs is smaller.

\subsection{Evolution of the physical properties of an accreting WD during a full Nova cycle}
\label{subsec:cycle}
\begin{figure} 
\includegraphics[width=\columnwidth]{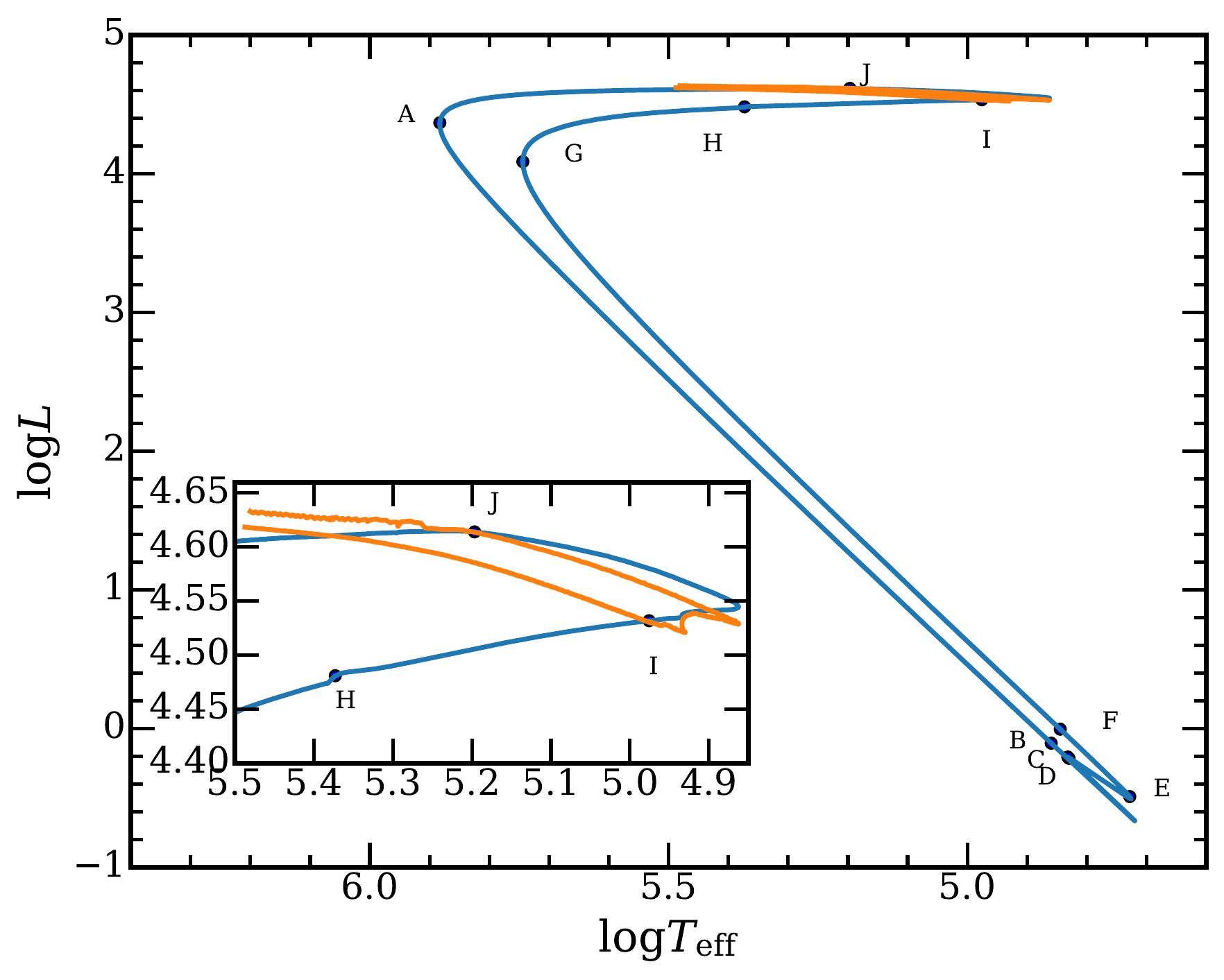}
\caption{
	Evolution  of the model with \mwd=1.20\msun, \mdot=$10^{-9}$ \msun\pyr\ and Z=$10^{-4}$ in the HR diagram for the no-mixing case. Along 
                 the track some relevant epochs are marked by capital letters: 
        \textbf{A}: bluest point along the loop; 
        \textbf{B}: minimum temperature of the H-burning shell; 
	\textbf{C}: full ignition of H-burning via CNO cycle; 
        \textbf{D}: flash-driven convection sets in; 
        \textbf{E}: convective shell attains the surface; 
        \textbf{F}: maximum luminosity of the H-burning shell;
        \textbf{G}: maximum temperature of the H-burning shell;
        \textbf{H}: convective shell recedes from the surface; 
        \textbf{I}: onset of mass loss via super-Eddington wind; 
        \textbf{J}: end of the mass loss episode.
Orange line shows the behaviour of $L_{Edd}$ close to and during mass-loss stage.}
    \label{hr_added}
\end{figure}

The evolution of H-accreting WDs via recurrent flashes has been addressed by many authors in the past. WDs are known to evolve along closed loops in the 
HR diagram, as displayed in Fig.~\ref{hr_added} where we report one nova outburst we computed for the model with CO WD mass \mwd=1.20 \msun, accretion rate 
\mdot=$10^{-9}$\,\myr and metallicity Z=$10^{-4}$ for the no-mixing case.
In this figure, several important epochs in the evolution along the loop in 
the HRD are annotated by letters. Physical properties of the WD at these
epochs are presented in Table~\ref{tab_added}. 

\begin{figure*}
	\includegraphics[width=\textwidth]{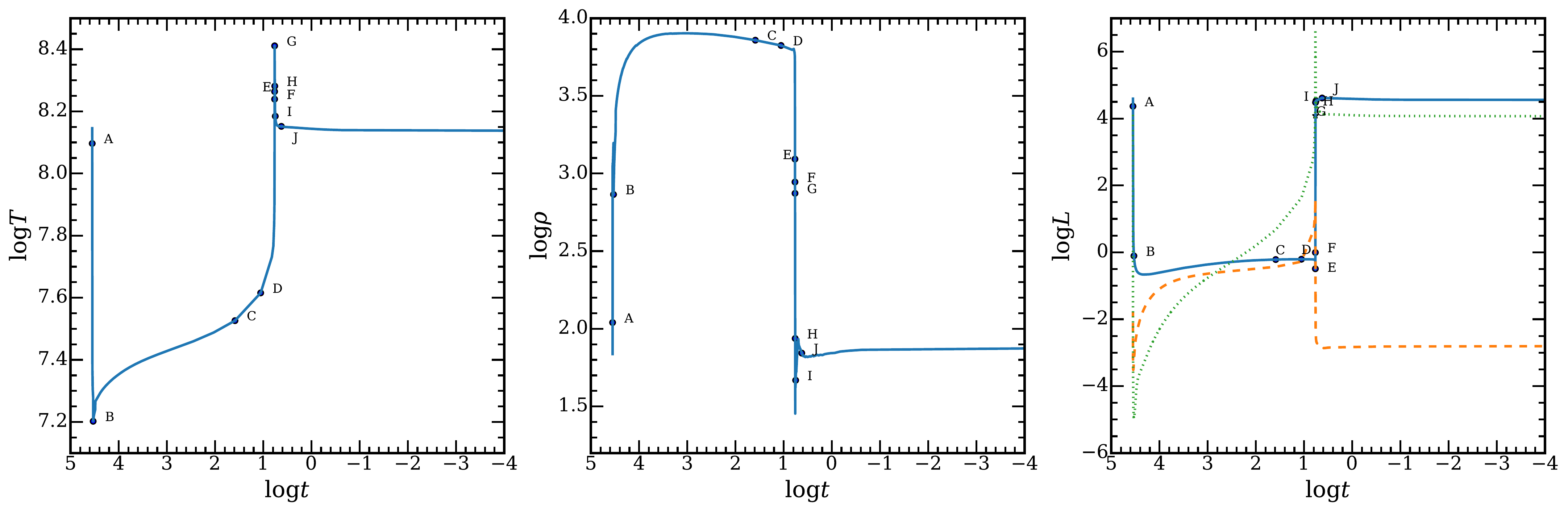}
	\caption{Time (remaining time of nova cycle) evolution of temperature (left panel) and density (middle panel) at H-burning shell for the same model as in Fig.~\ref{hr_added}. In the right panel 
                 we report the luminosity at the H-burning shell due to the $pp$-chain (dashed line) and CNO cycle (dotted line). For comparison we report also the total surface 
                 luminosity (solid line). Points and letters along the curves mark the same epochs displayed in Fig.~\ref{hr_added} and in Table~\ref{tab_added}. Time is in yr.
}
                 
    \label{hshell}
\end{figure*} 
\begin{table*}
	\centering
	\caption{Physical properties of the models with \mwd=1.20\msun, \mdot=$10^{-9}$\msun\pyr for the no-mixing 
        case. The letters in the first row refer to the epochs marked in Fig.~\ref{hr_added}. We list:
        $\Delta t$ -- the time elapsed from the previous epoch in yr (for epoch \textbf{A} it is 0);
        $\Delta M_{\rm acc}$ -- the amount of mass accreted from epoch \textbf{A} (in $10^{-5}$ \msun);
	$\log(L_{\rm H}/L_\odot)$ -- the H-burning luminosity at the H-burning shell; 
	$\rho_{\rm H}$ -- the density (in ${\rm g\cdot cm^{-3}}$) and $T_{\rm H}$ -- the temperature (in K) at the H-burning shell; 
        $\log(L/L_\odot)$ -- the surface luminosity; 
        $\log T_{\rm eff}$ -- the effective temperature (in  K).
        The upper part of the table refers to the models accreting matter with 
Z=$10^{-4}$, while the lower part --- to models with Z=$2\times 10^{-2}$.
}
	\label{tab_added}
	\begin{tabular}{llccccccccc} 
		\hline
		& {\bf A} & {\bf B} & {\bf C} & {\bf D} & {\bf E} & {\bf F} & {\bf G} & {\bf H} & {\bf I} & {\bf J} \\
		\hline
             \multicolumn{10}{c}{$Z=10^{-4}$} \\
		$\Delta t$ &0 & 1558.63 & 33900.45 & 27.34 & 5.50& 0.0006 & 0.0116 & 0.0405 & 0.135 & 1.423 \\
		$\Delta M_{acc}$ &0 & 0.156 & 3.546 & 3.548 & 3.549 & 3.549 & 3.549 &3.549 & 3.549 & 0.439 \\
		$\log(L_{\rm H}/L_{\odot})$ & 3.988 & -0.153 & -0.406 & 0.062 & 3.949 & 5.558 &5.464 & 3.847 & 3.532 & 4.123 \\
		$\log\rho_{\rm H}$ & 2.040 & 2.864 & 3.858 & 3.824 & 3.093 & 2.945 & 2.872 & 1.937 & 1.668 & 1.843 \\
		$\log T_{\rm H}$ & 8.096 & 7.202 & 7.526 & 7.615 & 8.263 & 8.239 & 8.410 & 8.281 & 8.184 & 8.151 \\
		$\log(L/L_\odot)$ & 4.366 & -0.107 & -0.215 & -0.206 & -0.491 & -0.005 & 4.085  & 4.480 & 4.531 & 4.614 \\
		$\log T_{\rm eff}$ & 5.882 & 4.859 & 4.829 & 4.831 & 4.728 & 4.844 & 5.743 &5.372 & 4.975 & 5.196 \\
             \multicolumn{10}{c}{$Z=2\times 10^{-2}$} \\
		$\Delta t$  & 0 & 941.18 & 18603.19 & 32.14 & 0.658 & 1.032e-5 & 6.520e-5 & 2.835e-4& 3.765e-4 & 0.360 \\
		$\Delta M_{acc}$ & 0 & 0.094 & 1.954 & 1.957 & 1.957 & 1.957 & 1.957 &1.957 & 1.957 & 0.043 \\
		$\log(L_{\rm H}/L_{\odot})$ & 4.07 & -0.86 & -0.74 & 1.11 & 6.64 & 6.84 & 5.52 & 6.236 & 5.83 & 4.12 \\
		$\log \rho_{\rm H}$ & 1.575 & 2.827 & 3.795 & 3.740 & 3.248 & 3.100 & 2.40 & 1.36 & 1.19 & 1.470 \\
		$\log T_{\rm H}$ & 7.985 & 7.098 & 7.375 & 7.542 & 8.105 & 8.146 & 8.372 & 8.226 & 8.186 & 8.013 \\
		$\log(L/L_\odot)$ & 4.45 & -0.68 & -0.54 & -0.53 & -0.69 & -0.72 & 3.795 & 4.458 & 4.47 & 4.57 \\
		$\log T_{\rm eff} $ & 5.93 & 4.72 & 4.75 & 4.75 & 4.70 & 4.69 & 5.71 & 5.32 & 5.15 & 5.90 \\
		\hline
	\end{tabular}
\end{table*}
At the epoch corresponding to the bluest point of the track (point \textbf{A} in Fig.~\ref{hr_added}),  the H-burning shell 
is no longer able to provide the energy to balance the radiative losses from the 
surface
\citep[see  discussion in ][]{iben82}, so that the luminosity of the WD 
decreases, the external layers contract and the model evolves along the 
cooling sequence. During the following phase, lasting for 
$\Delta t_{AB}= 1558.63$\;yr,  mass deposition determines 
the compressional heating of the H-rich mantle, even if the delivered energy is not able to counterbalance the radiative losses and the H-burning shell progressively cools 
down (see right panel in Fig.~\ref{hshell}). The energy contribution coming from the CNO cycle rapidly extinguishes, even if H-burning never completely dies, as $pp$ and 
$pep$ reactions are always active as well as $p$-capture on \isot{12}{C} (see right panel in Fig~\ref{hshell}). 
Due to the combined action of fresh matter accretion and contraction of the H-rich mantle, at the epoch \textbf{B} in Fig.~\ref{hr_added}, the H-burning shell ceases 
cooling down, while the local density continuously increases. As the H-rich layer grows in mass, the temperature of the H-burning shell $T_H$ progressively increases and, 
when it attains $\sim 3.7\times 10^7$~K after $\Delta t_{BC}= 33900.45$\;yr, the 
CNO cycle is fully active\footnote{We assume 
that CNO is fully active when its contribution to the total surface luminosity is 
10 times that due to the $pp$ chain. 
Note that at this epoch 
the mass fraction abundance of \isot{16}{O} at the H-burning shell has already been 
reduced by $\sim$5\%.}. 
Due to the partial degeneracy of the matter at the H-burning shell and the 
continuous mass deposition, the local energy production largely increases and after $\Delta t_{CD}= 27.34$\;yr convection sets in, 
because the timescale in which nuclear energy is delivered is shorter than the local thermal diffusion timescale, and in $\Delta t_{DE}= 5.50$\;yr the 
flash-driven convective shell attains the surface of the accreting WD. 
The resulting H-flash turns into a thermonuclear runaway which 
drives the evolution 
toward higher luminosity and larger effective temperature until, after 
$\Delta t_{EF}= 5.26$\;hr the H-burning luminosity attains a maximum,
though the temperature in the burning shell still continues to rise for
$\Delta t_{FG}=4.24$\;day. 
When the model attains a surface luminosity of $\sim 10^{4}$\lsun, it evolves redward at almost constant luminosity. 
After 
$\Delta t_{GH}= 14.8$\;day, convection starts to recede from the surface, 
but the star still expands, evolving redward. During this phase, the surface luminosity becomes 
larger than the Eddington limit after $\Delta t_{HI}=49.3$\;day, and mass loss via 
stellar wind begins.

When the large energy excess produced by the thermonuclear runaway has been dissipated via both mechanical work to expand the H-rich mantle and stellar wind mass 
loss, the H-burning shell moves outward in mass rapidly converting hydrogen into helium. 
The reduction of the H-rich mantle via mass loss and nuclear burning 
determines the blue-ward evolution of the model up to when, after 
$\Delta t_{IJ}=1.423$\;yr, mass loss ceases. 
It is worth noticing that during the mass loss phase the Eddington luminosity 
progressively increases, due to a decrease of the average opacity in the envelope 
$\overline{\kappa}$ (see the inside panel in Fig.~\ref{hr_added}). At epoch \textbf{J} the energy production via H-burning 
almost completely counterbalances the radiative losses from the surface; the additional energy contribution comes from the contraction of the H-rich mantle. 
As the star consumes hydrogen, the energetic contribution of nuclear burning decreases and that of the contraction increases. The latter determines the blue-ward evolution of the WD. When the gravitational energy release exceeds $\sim$5\% of the total released energy, the star attains the bluest point in the HR diagram and H-burning rapidly declines.

This general scenario is valid also for high metallicity models, even if some important differences exist. First, as the CNO abundance in the accreted matter
is larger, H-burning is ignited sooner, at lower $T_H$; this implies that a lower mass has to be accreted in order to trigger the TNR and, hence, the degeneracy level at the 
H-burning shell is lower (slightly lower ignition density -- see Table~\ref{tab_added}). It has to be remarked also that, due to the larger CNO abundance in the $Z=2\times 10^{-2}$ 
model, the CNO cycle provides an energy contribution comparable to that from the $pp$-chain. 
The reduced degeneracy at the H-burning shell and the larger CNO abundance in high metallicity models act in opposite directions, the latter causing a stronger 
H-flash, the former allowing a larger outward flux of thermal energy produced during the H-flash and, hence, a less rapid increase of the local temperature. We found that in the Z=$2\times 10^{-2}$ case $L_H^{max}$ is larger while $T_H^{max}$ is lower as compared to the Z=$10^{-4}$ one 
(see Table~\ref{tab_added}). Notwithstanding, the maximum surface luminosity during the loop is quite similar, while 
the percentage of the mass transferred during the full nova cycle effectively retained by the accreting WD is 
in the former model $\sim 2\%$ and in the latter $\sim 12\%$, respectively. 
Such an occurrence is, once again, a direct 
consequence of the different metallicity, the Z=0.02 model having a larger thermal content per unit mass in the H-rich mantle. In addition, the surface opacity 
is also larger so that the corresponding Eddington luminosity is lower and, hence, in our computations mass loss starts when the model is more compact with respect 
to the $Z=10^{-4}$ case (see Tab.~\ref{tab_added}). This implies that in the high metallicity model a very small amount of the energy delivered via nuclear burning is employed to expand the model 
before the onset of super-Eddington wind mass loss and, hence, a larger portion of the accreted matter is expected to be lost. 

Another important aspect related to the metal content (i.e. CNO abundance) of the accreted matter is the abundance of the $\beta$-unstable isotopes 
\isot{13}{N}, $^{14,15}{\rm O}$, and \isot{17}{F} in the convective envelope of the accreting WD. These isotopes are produced by the hot CNO cycles and, since they 
have a longer $p$-capture timescale with respect to the convective mixing timescale, they are carried outward by convection. Then, their local decay into their daughter 
nuclei ${\rm {^{13}C}}$, ${\rm {^{15}N}}$, and ${\rm {^{17}O}}$ provides additional power to trigger the expansion and the ejection phase. In our models, as the maximum 
attained temperatures during the outbursts are quite similar (see Table~\ref{tab_added}), the amount of $\beta$-unstable isotopes dredged up to the surface scales as the 
total metallicity, being larger at Z=0.02.

\begin{figure} 
    \includegraphics[width=\columnwidth]{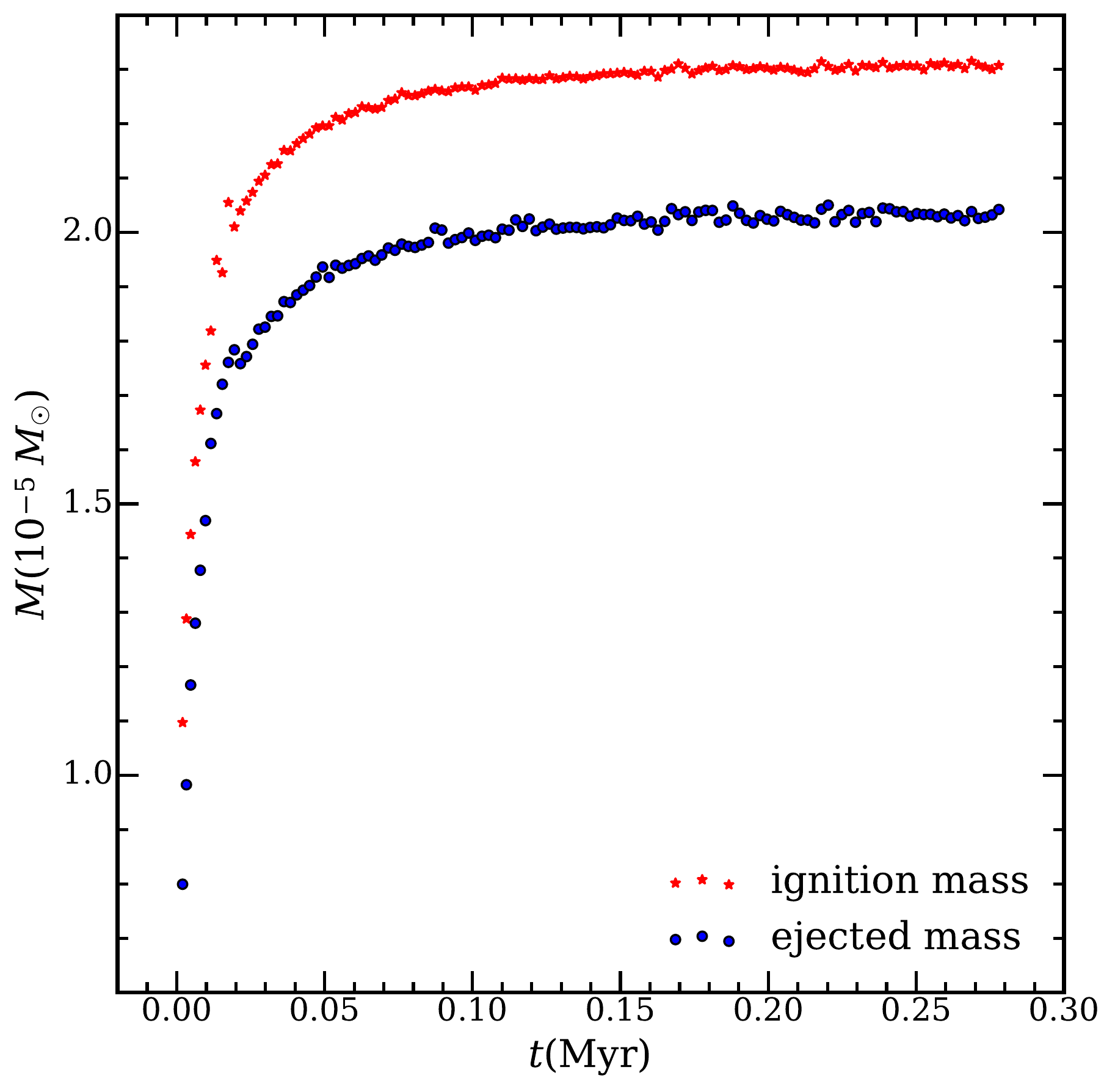}
	\caption{ Evolution of ignition mass  
and ejected mass  
as a function of time from the onset of mass transfer                                          
for accreting WDs with \mwd$= 1.00\;{\rm M_{\odot}}$,
accretion rate \mdota$= 10^{-8}\;{\rm M_{\odot}/yr}$, metallicity $Z = 0.02$ and without mixing.
}
    \label{fig:nova_pro_diff_outburst}
\end{figure}

In Fig.~\ref{fig:nova_pro_diff_outburst}, we show the evolution of 
ignition mass and  ejected mass 
for multiple nova outbursts in the initial stage of our calculations, while
models attain a steady-state pattern of outbursts, for \mwd$= 1.00\;{\rm M_{\odot}}$, accretion rate \mdota$= 10^{-8}\;{\rm M_{\odot}/yr}$, $Z = 0.02$ and no mixing. Ignition mass and ejected mass are defined as the amount of matter accreted and ejected during one full nova cycle, respectively (see above). 
In the following, we consider the properties of the last computed flash as typical for novae with given \mwd\ and \dotm.

\begin{figure*} 
\includegraphics[width=\textwidth]{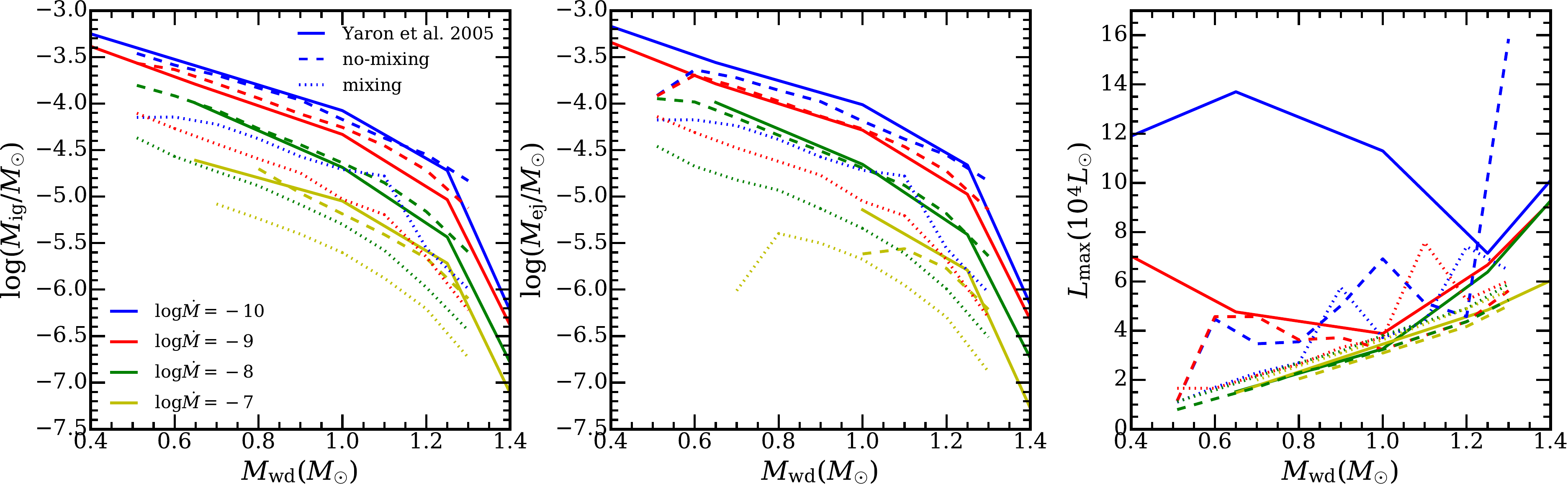}
\caption{Comparison of ignition mass (left panel), ejected mass (middle panel) and maximum luminosity (right 
panel) for our models with $Z = 0.02$
	with the results of \citet{ypsk05}. 
Note that convergence problems did not allow 
	to compute enough outburst cycles for models with \mdota$=10^{-10}\,{\rm M_{\odot}/yr}$, leading to a large scatter in the right panel.
 Note, the models of  \mwd=0.8, 0.9\,\msun accreting at the rate
 $10^{-7}$\,\myr\ in no-mixing case do not eject matter.                                        
}
\label{fig:nova_com_yaron}
\end{figure*}

\subsection{Novae properties at metallicity $Z = 0.02$ and $10^{-4}$}
\label{novaproperties}

In Tables~\ref{tab:nova_z2m2} and \ref{tab:nova_z1m4}, we present
characteristics of novae outbursts for the models with $Z= 0.02$ and $Z=
10^{-4}$ without mixing of accreted matter: accreted mass $M_{\rm acc}$, ejected
mass \mej,  and the maximum luminosity during
the outburst $L_{\rm max}$. 
The results for the models with mixing are presented in 
Tables~\ref{tab:nova_z2m2_mx} and \ref{tab:nova_z1m4_mx}.
Recall that in the computations of these models the WDs accrete matter with enhanced C, O and Ne abundance (see \ref{subsec:nova_cal}). 
Physically, this procedure corresponds to the assumption that 
at each time step the accreted matter $\Delta M_{acc}$, with Z being that of the donor, is instantaneously 
mixed with a fraction of the underlying C- and O-rich core corresponding 
to $\Delta M_{mix}={\frac{1}{3}}\cdot\Delta M_{acc}$. As a consequence, the accreted 
matter 
progressively penetrates the underlying WD, which becomes gradually eroded. 
Thus, for the models with mixing and given a mixing fraction of 0.25, at the instant of
ignition the total mass of the C- and O-enriched layer is

\begin{equation}
	M_{ign}^{tot}=\int_{loop} (\dot{M}_{mix}+ \dot{M}_{acc}) dt= \int_{loop} \dot{M}_{acc} (1+\frac{1}{3}) dt=\frac{4}{3}M_{acc}.
\label{eq:mign}
\end{equation}
Equation~(\ref{eq:mign}) implies that in our calculations of the mixing case, the conditions for ignition are ultimately set by the amount of matter transferred from the donor, as in the no-mixing case.
 
In Fig.~\ref{fig:nova_com_yaron} we compare our models with $Z = 0.02$ and the 
models of \citet{ypsk05}.
In making such a comparison, it is important to keep in mind the different numerical implementations, particularly the different prescriptions invoked for mass loss and for diffusion at the core/envelope interface, employed by \citet{ypsk05}.\footnote{For more details on their numerical method, see also \citet{pk95}.} Furthermore, in comparing 
\citet{ypsk05} and our results it should be noted that while we take the "last" 
outburst as typical, \citet{ypsk05} present characteristics of a random outburst; therefore a direct comparison may be misleading
if in either case the steady state is not reached.
Nevertheless, the values of \mign\ and \mej\ obtained in the  no-mixing case for the 
\mwd$\leq1$\,\msun\ and $10^{-10}\leq \mdot \leq10^{-8}$\,\myr\ models reasonably agree with the values from 
\citet{ypsk05}. This may mean that in this range of WD masses the timescale of diffusion exceeds (1-10)\,Myr
and it does not influence the hydrogen ignition process.  
Intersection of lines for these parameters may be attributed to numerical 
noise. 
We cannot draw a definitive conclusion with respect to the behaviour of 
$L_{\rm max}$ for 
$\dot{M} \leq 10^{-9}$\,\msun/yr, since in our calculation, due to convergence problems, the steady-state was not reached. 
However, we note that the non-monotonic behaviour for $L_{\rm max}$ found by \citet{ypsk05} for  low 
\dotm\ models may  hint to the same circumstance. Generally, we would expect that, with decreasing \mdota,
the maximum luminosity would increase, since H-ignition occurs in more degenerate matter.

Compared to the models without mixing, in the models with mixing 
ignition masses, ejected masses, and recurrence periods\footnote{The recurrence period $P_{rec}$ is the time elapsing between two successive 
epochs of maximum luminosity. It can be estimated as the ratio of the mass accreted during a full nova cycle over the accretion rate.} 
are lower, as expected for a higher initial abundance of \isot{12}{C}, 
if mixing before the TNR is assumed.
For the same reason, the maximum luminosity is 
larger in these models (as long as we consider steady-state models). We note a relatively large difference 
between our results for mixing models and the results of Yaron et al., which is 
due to the larger degree of enrichment of the accreted layer by \isot{12}{C} than may 
be enabled by diffusion.

Notably, among our models there are some with \mej=0.
Similar models have been found previously as well. 
In these models, the H-shell does not expand significantly during the outburst, leading to small photospheric radii and high effective temperatures. 
Therefore, these objects are likely to be bright in the extreme ultraviolet and supersoft X-rays, but faint in the optical band and difficult to observe 
\citep{sps77,fuji82,ritt90}. 
The supersoft X-ray source ASASSN-16oh was suggested as 
the first candidate for this kind of object
\citep{2019ApJ...879L...5H}, but  \citet{2019arXiv190702130M} advanced an alternative 
interpretation, associated with the features of accretion onto WD for
this X-ray source.


\section{Discussion}   
\label{sec:disc}
\subsection{Influence of metallicity upon nova outbursts}

\begin{figure*}  
    \includegraphics[width=\textwidth]{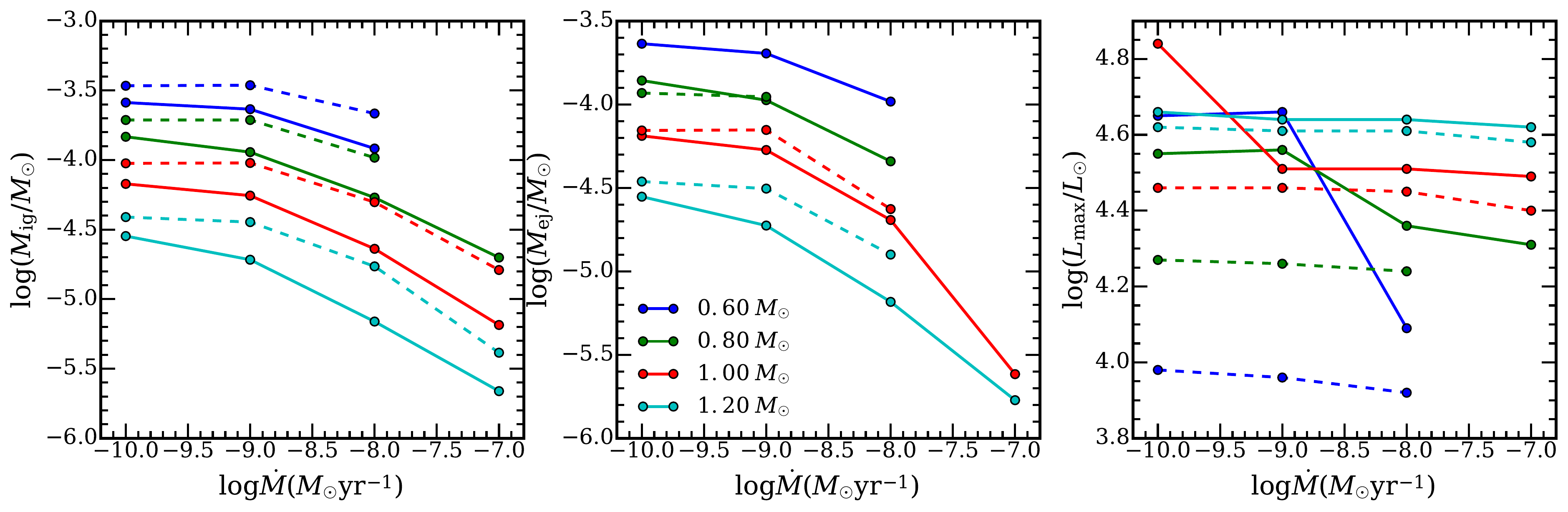}
    \caption{Comparison of novae properties for models without mixing, and metallicities $Z = 0.02$ (solid lines) and $Z = 10^{-4}$ (dashed lines). 
The lines for low mass WDs truncate at $10^{-8}\;{\rm M_{\odot}}/{\rm yr}$, since
the lower boundary of the steady burning regime of these white dwarfs is located between $10^{-8}\;{\rm M_{\odot}}/{\rm yr}$ and $10^{-7}\;{\rm M_{\odot}}/{\rm yr}$.
    In the middle panel, the models with \mej=0 are not plotted.
(For \mwd=0.6\,\msun\ and \mdot=$10^{-9}$ and $10^{-10}$\,\msun/yr the models did not reach a steady state and $L_{max}$ is not well defined).   
  }
    \label{fig:nova_com_diff_Z}
\end{figure*}

\begin{figure*}  
    \includegraphics[width=\textwidth]{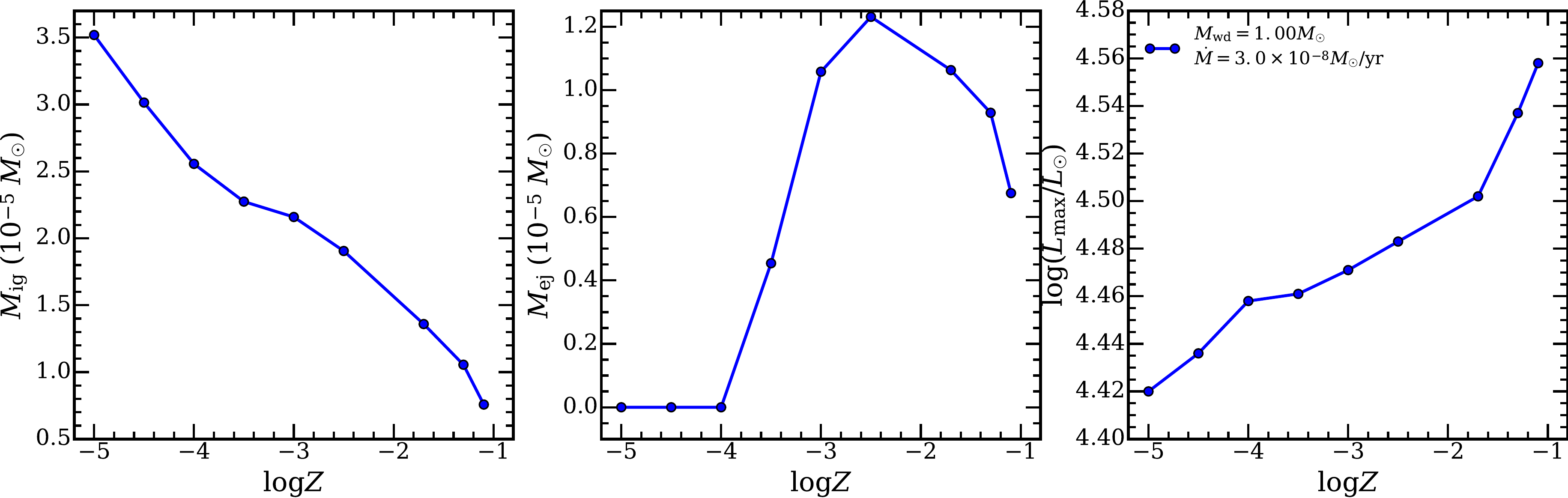}
    \caption{Dependence of nova properties on the metallicity for the white dwarf mass
    \mwd$= 1.00\;{\rm M_{\odot}}$ and accretion rate $\dot{M}=3.0\cdot10^{-8}\;{\rm M_{\odot}}/{\rm yr}$.
    In these calculations, we do not take mixing into account. The initial abundance of accreted matter is 
    computed as we describe in Sec.~\ref{subsec:nova_cal}.}
    \label{fig:nova_mul_Z}
\end{figure*}

Figure~\ref{fig:nova_com_diff_Z} presents a comparison of the properties of novae with
$Z = 0.02$ and $Z = 10^{-4}$. A reduced abundance
of CNO-elements results in a decrease in the opacity of the envelope matter,
allowing more heat to leak out of compressed layers faster, 
leading to a slower increase of the temperature.   As a result, 
more mass must be accreted prior to a TNR. The lower abundance of CNO-isotopes is 
balanced by a higher degree of degeneracy of the matter at the base of the H-shell and, hence, stronger outbursts and an even larger or comparable amount of ejected mass. However, 
this difference
decreases with decreasing mass. In models with \mwd$= 0.51\;{\rm M_
{\odot}}$ and \mwd$= 0.60\;{\rm M_{\odot}}$ and accretion rates $\dot{M}=10^
{-10}$ and $10^{-9}$\,\myr, the luminosity during nova outbursts never exceeds the
Eddington luminosity and \mej=0
(Table~\ref{tab:nova_z1m4}).
Evidently, for similar values of \mdota, the recurrence periods of models with low Z are larger than for models with high Z.

In order to  understand the dependence of nova properties on metallicity more comprehensively, we computed the 
evolution of accreting WDs for 
\mwd$= 1.00\;{\rm M_{\odot}}$ and $\dot{M}=3.0\cdot10^{-8}\;{\rm M_{\odot}}/{\rm yr}$ for a more detailed grid 
of $Z$ between $10^{-5}$ and $0.08$. 
 For these models, we have computed a large number of outburst cycles in order to 
be sure that a regular cyclic pattern is obtained. 
Figure~\ref{fig:nova_mul_Z} shows the dependence of novae properties on metallicity.
It is clear that accreting WDs with smaller metallicities have 
larger ignition mass and smaller maximum luminosity, as explained above.
However, the dependence of \mej\ on $Z$ is not monotonic. 
For $Z \lesssim 0.003$, the ejected mass becomes larger for higher metallicity. 
The reason is the following: on one hand, the nuclear luminosity increases as the metallicity increases, hence 
the ejected mass increases as the metallicity increases if $Z \lesssim 0.003$;
on the other hand, mass loss occurs at smaller photospheric radii as the luminosity 
increases. Hence, the escape velocity becomes larger, leading to the decrease of \mej.  
For the lowest $Z$, luminosity never exceeds \ledd\ and by virtue of 
Eq.~(\ref{eq:wind_rate}), outbursts do not result in mass-loss.

\subsection{Influence of mixing process}

\begin{figure*}   
\includegraphics[width=2.1\columnwidth]{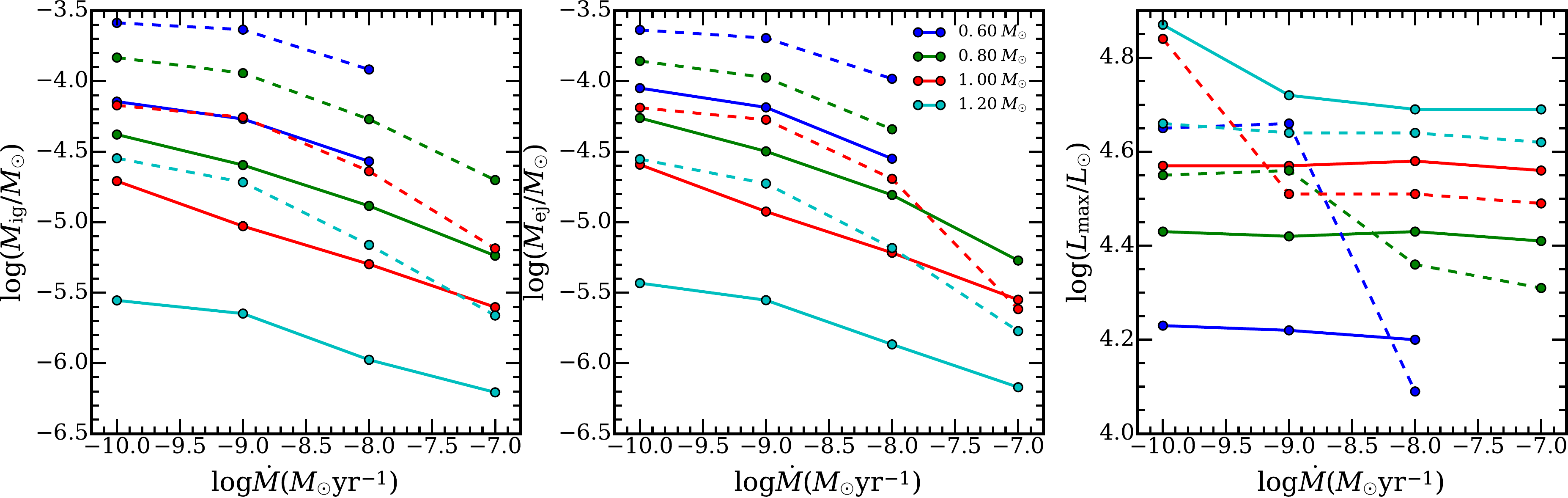}
\caption{Comparison of models with $Z = 0.02$ with mixing (solid line) and 
without it (dashed line).   
}
\label{fig:nova_com_diff_mx}
\end{figure*}

In Fig.~\ref{fig:nova_com_diff_mx}, the properties of models with and without mixing
are compared.
In the models which include our ``mixing'' approximation, 
the metallicity of the H-shell is greater, therefore
the differences between the models with and without mixing are similar to the differences between models 
with high and low metallicity. 
Thus, for the same reasons as presented in the 
previous subsection, 
\mign, \mej, and $P_{rec}$ 
are smaller in the models which include mixing than in the models without mixing,
 and the maximum luminosity is larger in the models with mixing.
Note that the behaviour of $L_{\rm max}$ in models with mixing (the rightmost panel of 
Fig.~\ref{fig:nova_com_diff_mx}) follows an irregular pattern. While the origin of this behaviour is unclear at 
present, we speculate that it may arise due to the interplay between the rate of energy release, the rate of 
energy transfer
to the surface by convection for outbursts of different strength, and the implemented mass-loss algorithm.

\subsection{Dependence of retention efficiency on metallicity and mixing}

\begin{figure}  
    \includegraphics[width=\columnwidth]{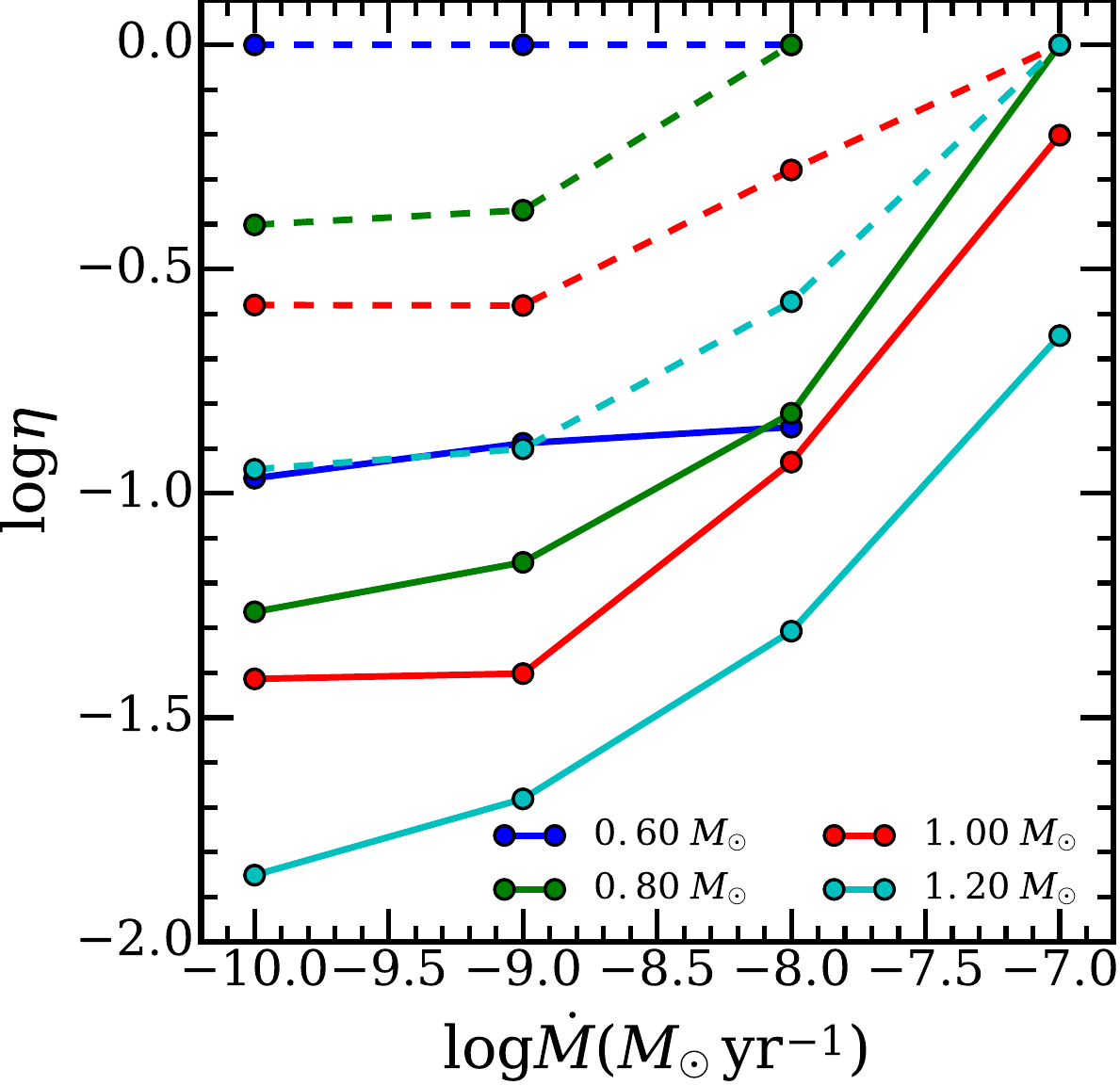}
    \caption{Comparison of retention efficiency for models with $Z = 0.02$ (solid line) and $10^{-4}$ (dashed line), without mixing. 
In the models with $Z = 10^{-4}$ and \mwd$=0.60\;{\rm M_{\odot}}$ 
luminosity during
    novae outbursts never exceeds
Eddington luminosity. 
  }
    \label{fig:eff_com_diff_z}
\end{figure}

\begin{figure} 
    \includegraphics[width=\columnwidth]{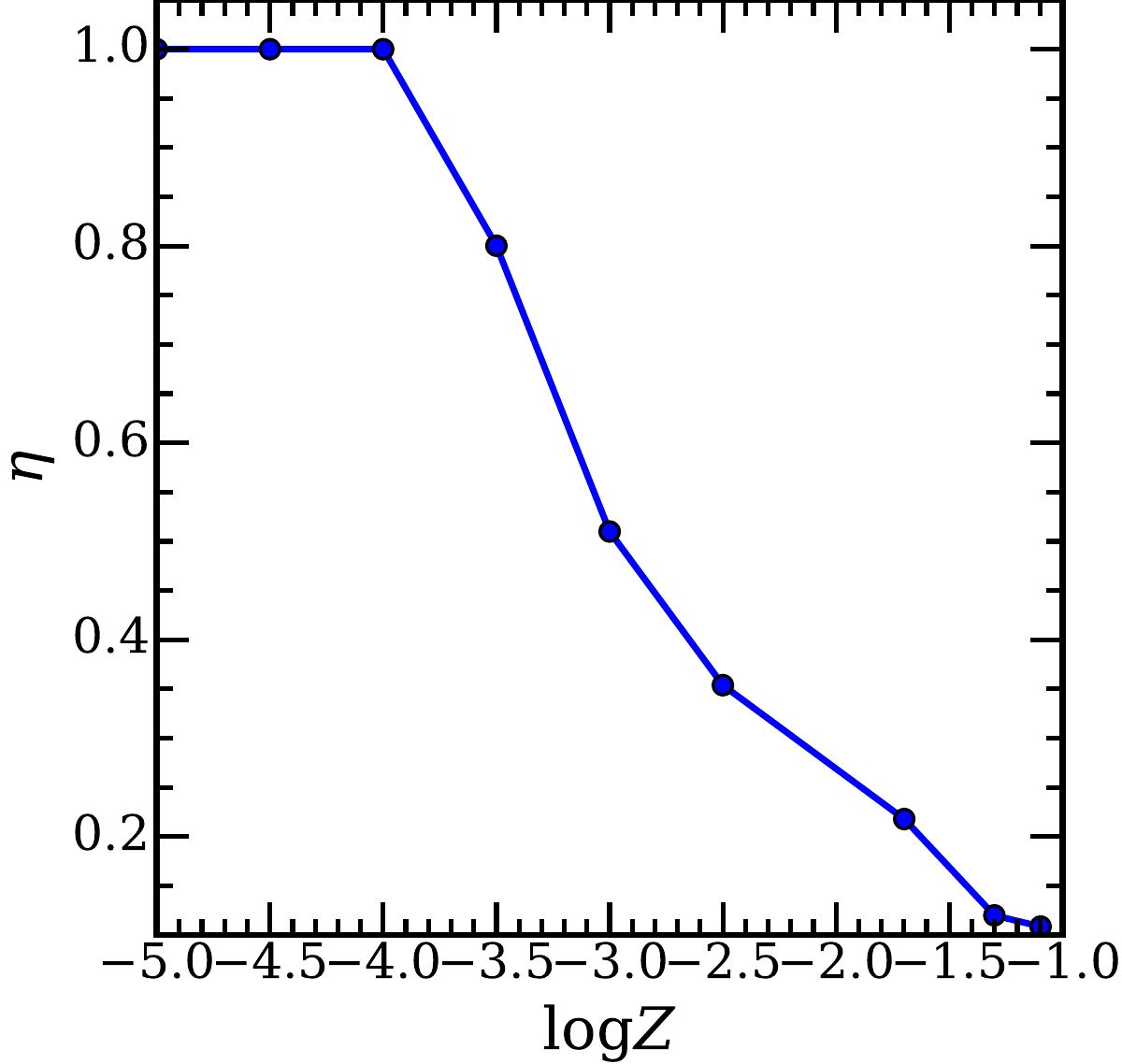}
    \caption{Dependence of retention efficiency on metallicity for $1.0\;{\rm M_{\odot}}$ WD with accretion rate $\dot{M} = 3.0\cdot10^{-8}\;M_{\odot}/{\rm yr}$.
    In these calculations, we do not take mixing into account.
    }
    \label{fig:eff_com_mul_z}
\end{figure}
\begin{figure}
    \includegraphics[width=\columnwidth]{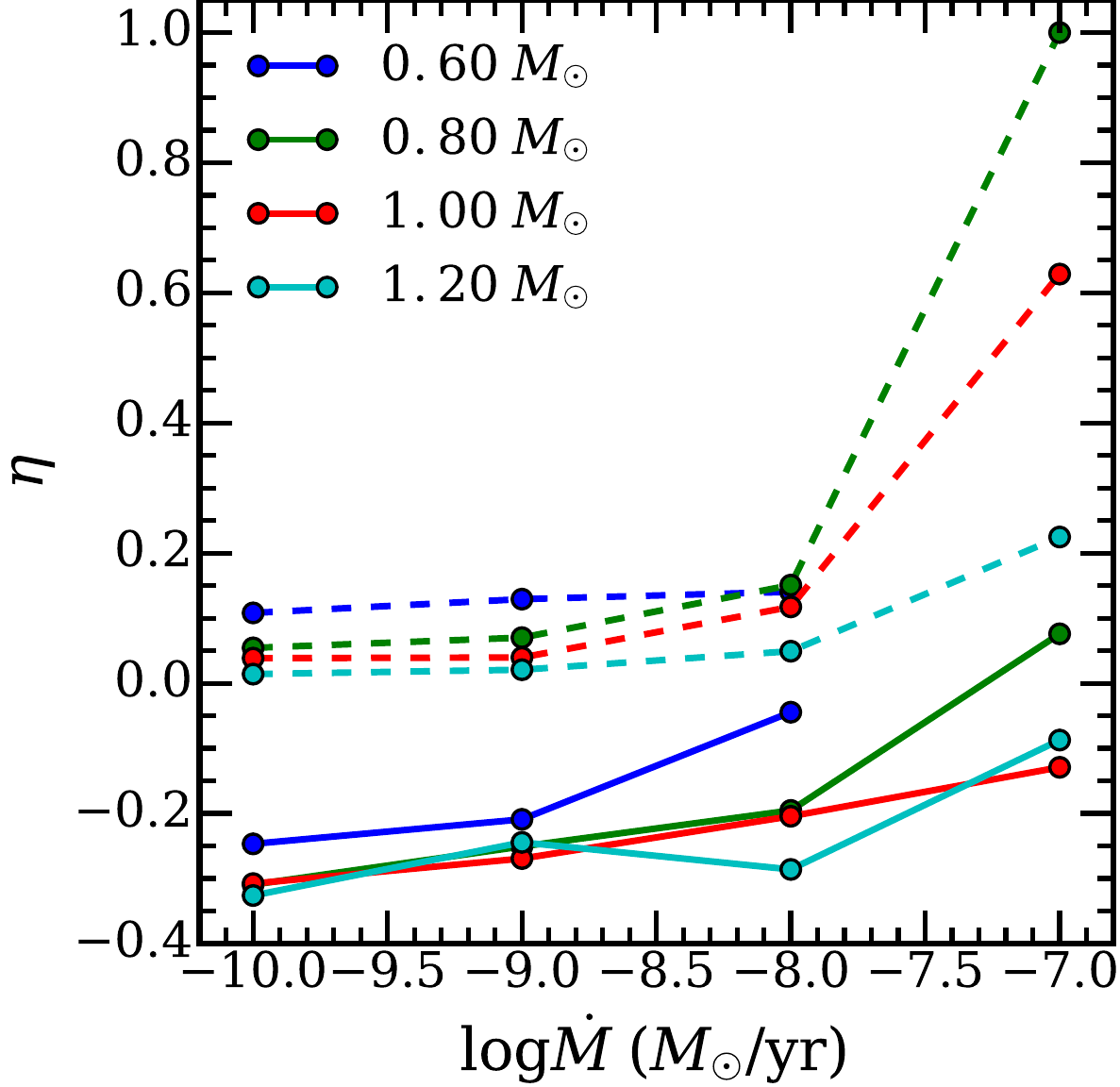}
    \caption{Comparison of retention efficiency for models with $Z = 0.02$, mixing (solid line) and no mixing (dashed line). 
}
    \label{fig:eff_com_diff_mx}
\end{figure}
\begin{figure} 
    \includegraphics[width=0.90\columnwidth]{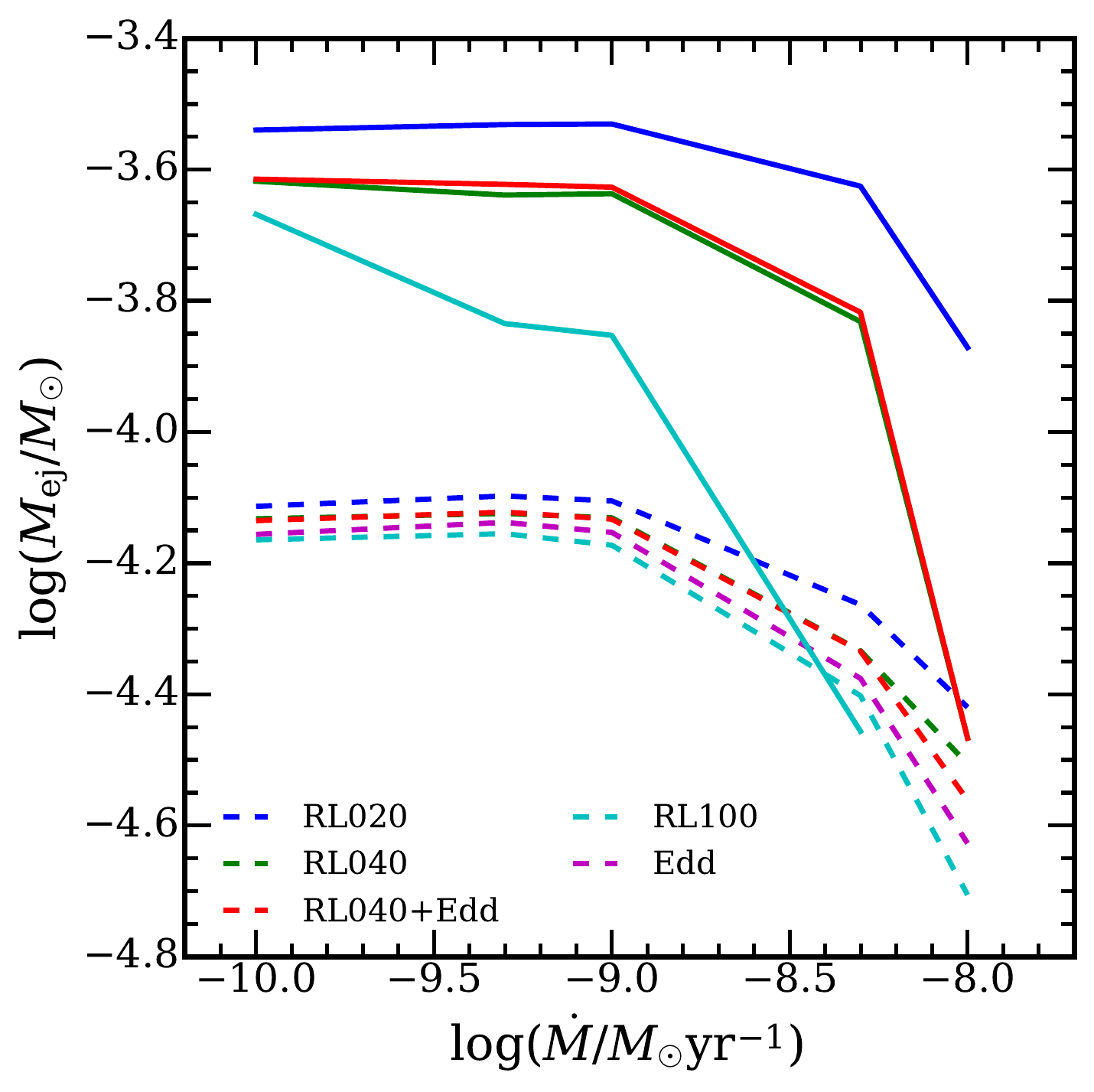}
    \includegraphics[width=0.90\columnwidth]{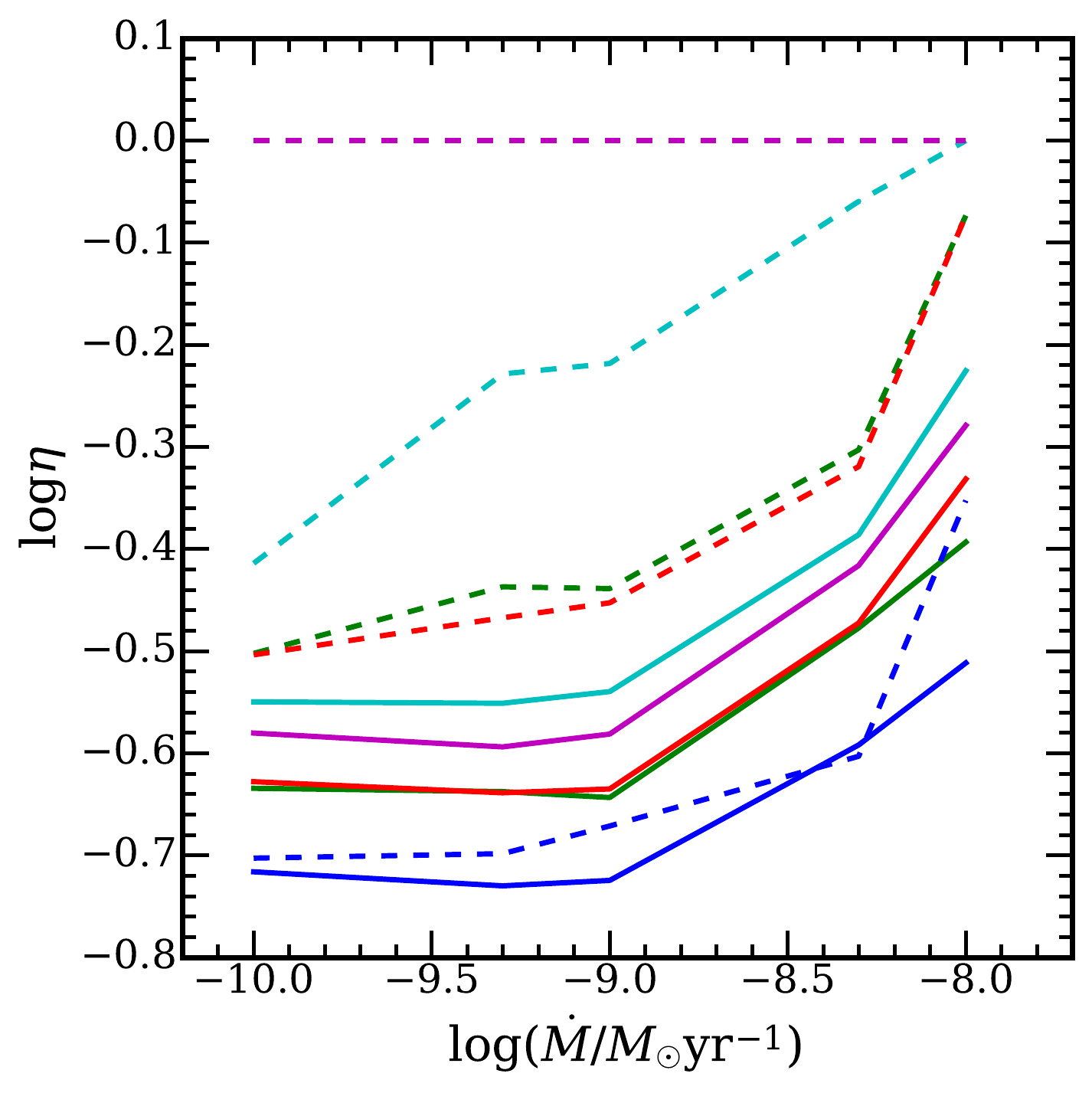}
	\caption{Comparison of ejected mass (upper panel) and retention efficiency (lower panel) for an accreting $0.60\;{\rm M_{\odot}}$ (dashed lines) and $1.0\;{\rm M_{\odot}}$ (solid lines) WD with different mass loss prescriptions, i.e. super-Eddington wind and Roche lobe overflow at $Z = 10^{-4}$ without mixing. 
Three constant values of Roche lobe radii were assumed: 0.2, 0.4, and 1.0\,\rsun.  
The model with \mwd$= 0.60\;M_{\odot}$ under assumption of super-Eddington wind only does not eject mass and it is not plotted in the upper panel.
}
\label{fig:para_diff_ml}
\end{figure}
In Fig.~\ref{fig:eff_com_diff_z}, we show the retention efficiency (Eq.~(\ref{eq:reten})) found for different WD masses, accretion rates, and metallicities. 
We find that the retention efficiency is higher in models with lower metallicity, less massive WDs or higher accretion rates. For more massive WDs or low accretion rates, the degeneracy in the H-layer is higher. For higher metallicity,
the nuclear energy production is higher. These circumstances lead to more violent outbursts and lower retention efficiency.  

Figure~\ref{fig:eff_com_mul_z} demonstrates the retention efficiency found for our models of a $1.0\;{\rm M_{\odot}}$ WD with accretion rate 
$\dot{M} = 3.0\times10^{-8}\;{\rm M_{\odot}/yr}$ and different metallicities.
This plot clearly shows that retention efficiency decreases with increasing metallicity.

Figure~\ref{fig:eff_com_diff_mx} shows the retention efficiency for WDs with 
different masses, with and without mixing. 
Note that the solid line for $1.20\;M_{\odot}$ in the figure intersects other lines, due to numerical noise.
We find that the retention efficiency varies for different outbursts and the novae properties do not converge to an 
asymptotic limit for our models with the lowest accretion rates.
In the mixing case the retention efficiency is mostly negative because, as discussed in \S~\ref{novaproperties} (see Eq.~(\ref{eq:mign})), the mass involved in the 
TNR is a factor 4/3 larger than the mass accreted during a full 
nova cycle $M_{acc}$. 
This implies that, except in models with high accretion rates, the C/O enhancement in the accreted layers determines a secular
reduction of the underlying WD mass.
When comparing our results with those by \citet{ypsk05}, we find that for low values of \mdot\; our estimated retention efficiencies are 
lower. This indicates that the C/O enrichment we adopted is definitively larger than that enabled by diffusion.

\subsection{Influence of different mass loss algorithms}
\label{subsec:inf_ml}

As discussed above, there exist several algorithms in the literature for prescribing mass loss in hydrostatic models of outbursts from accreting WDs. 
In order to evaluate the effect of different mass loss prescriptions on our results, we carried out trial computations using an alternative simple model in which it is assumed that expanding WD experiences Roche lobe overflow. We computed models of  $0.60\;{\rm M_{\odot}}$ and $1.0\;{\rm M_{\odot}}$ WDs  with $Z= 10^{-4}$ which accrete 
un-enriched matter at different rates. Roche lobe radii were set to 0.2, 0.4, and 1.0\,\rsun. Necessarily, these models 
are not entirely self-consistent, e.g., in any given cataclysmic variable, the mass of the WD, Roche lobe radius, and accretion rate are interrelated via the evolutionary state of the binary, while in symbiotic binaries the notion of a Roche lobe radius is hardly applicable to outbursting WDs. These results are therefore only intended for a qualitative comparison with our models.

The results of these trial computations are presented in Fig.~\ref{fig:para_diff_ml}.
For \mwd=0.6\,\msun, WD outbursts never result in mass ejection if the SEW algorithm
is assumed. For the RLOF algorithm, however, we find that outbursts may result in the loss of up to (70-80)\% of the accreted mass, depending on their strength. Comparing Figs.~\ref{fig:eff_com_diff_z} and \ref{fig:para_diff_ml}, we see that for 
\mwd=1\,\msun\  the retention efficiency
in the SEW case may be slightly higher than in RLOF case, indicating that SEWs
remove most of the mass before RLOF. A similar relation between these two algorithms 
was demonstrated by \citet{wbbp13} for WDs accreting solar metallicity matter. 
This inference is confirmed by a computation of a sequence of models
for 1\,\msun\ WD, in which both SEWs and RLOF were taken into account (for a Roche lobe radius of 0.4\,\rsun): in this particular case, the difference in 
$\log(\eta)$ did not exceed 0.04dex. From Figure~\ref{fig:para_diff_ml}, however, we may conclude that differing assumptions made in the literature regarding the dominant mechanism driving mass loss in nova eruptions may result in retention efficiencies differing by up to factor $\lesssim 4$, depending on the mass of the accretor, the mass transfer rate, and the dimensions of the system. In light of this analysis, it is clear that the mass-loss algorithm assumed in hydrostatic models remains a critical uncertain parameter in modelling nova eruptions and populations of cataclysmic variables. 

This conclusion is confirmed by results of \citet{2017hsn..book.1211S}, who
compared retention efficiency $\eta$ for 1.35\msun\  WD
accreting solar composition matter at a rate of $1.6\times10^{-9}$\,\myr\ assuming three
different mass loss prescriptions: SEW,  RLOF and ejection of
matter that exceeded the escape  velocity and became optically thin
\citep{2009ApJ...692.1532S} .
They show that the values of $\eta$ associated with SEW and RLOF mechanism are quite compatible, while the latter mechanism provides  retention efficiency higher by a large factor (up to $\sim 6$).

\subsection{Implications for populations of accreting WDs and SNe~Ia}

In modelling populations of  SNe~Ia progenitors in the single-degenerate (SD)
scenario, among the key ingredients  are the location and the breadth of the stability strip in the 
$M_{WD}-\mdot$ plane and efficiency of mass retention by 
accreting WDs in the unstable nuclear burning regime. Significantly differing results have been obtained by different authors, however, primarily due to differing 
prescriptions for enrichment and mass-loss. As a result, the delay-time distributions and rates found for SNe~Ia in the SD-scenario
vary substantially, even for solar Z \citep[see, e.g.,][]{btn13}.

In our analysis above, we found that location of the boundary of stable nuclear burning and the retention efficiency strongly depend on composition, thus influencing the rate of SNe Ia in the SD-scenario in stellar populations with different metallicities.
Additionally, we found that the retention efficiency can be negative for a wide range of accretion rates and WD masses, when enrichment of the accreted matter via mixing with the underlying WD
is taken into account. Leaving aside outstanding uncertainties in the physical mechanism(s) driving mass-loss (as discussed above), we note that metallicity may have an important effect not only for hypothetical precursors of SNe~Ia,
but also for models of other  
populations of accreting WD binaries, such as cataclysmic variables and symbiotic binaries.
These speculations must be verified by future binary population synthesis studies. 

\section{Summary and Conclusion}
\label{sec:sum}

Using the publicly-available stellar evolution code \textsc{mesa}, we modelled the evolution of accreting WDs with metallicity $Z = 0.02$ and $Z = 10^{-4}$. White dwarf  masses ranged from $0.51\;{\rm M_{\odot}}$ to $1.30\;{\rm M_{\odot}}$ and accretion rates ranged from $10^{-10}\;{\rm M_{\odot}}/{\rm yr}$ to $10^{-6}\;{\rm M_{\odot}}/{\rm
yr}$. For each model, we computed the evolution of accreting WDs with and
without a simplified approximation for mixing accreted matter with C/O from the underlying WD. We have investigated the properties of novae
in different models as well as the influence of metallicity and mixing on them.
 We use a mixing fraction of 0.25, close to the upper limit suggested by observations, in order to test the maximum effect that mixing may have upon novae explosions.
The main results are as follows:

1) We calculated the range  of accretion rates
allowing  stable H-burning for WDs with $Z =
0.02$ and $Z = 10^{-4}$ (see Fig.~\ref{fig:snbb_diff_Z}).
For $Z = 10^{-4}$ these rates are lower compared to $Z = 0.02$.
    
2) For both values of metallicity we computed the key properties of novae (i.e. ignition mass, ejected mass, 
maximum luminosity) for a comprehensive grid of accreting WDs 
with different WD masses, accretion rates and metallicities, 
see 
Tables~\ref{tab:nova_z2m2},~\ref{tab:nova_z1m4}, \ref{tab:nova_z2m2_mx},\ref{tab:nova_z1m4_mx}.

3) We confirm that metallicity has an important impact on the properties of novae. 
For models with $Z = 0.02$, the ignition masses and recurrence periods are smaller, 
while the ejected mass and maximum luminosity are larger in contrast to models with $Z = 10^{-4}$.
We find that retention efficiency during novae outbursts decreases with increasing metallicity.                                 

4) We remark that in mixing models, by virtue of Eq.~(\ref{eq:mign}), at the onset of
the first TNR, abundance of metals in the mixed envelope of WD is 
$$Z_{env}=(1/3 M_{acc} Z_{WD} + M_{acc} Z_\odot)/M_{ign}=0.265,$$
while $X_{env}=0.525$, $Y_{env}=0.21$.
Thus, after completion of an outburst cycle, if the ejected mass $M_{\rm ej}$ is lower than $M_{\rm ign}$
as defined in Eq.~\ref{eq:mign},
the surface of the WD should be represented by a He-rich layer having $Y=0.735$ and
$Z=0.265$, the latter including CNO isotopes, with relative abundances determined by
H-burning during the TNR, and original scaled solar mass fraction of
other metals. After the beginning of a new cycle, accreted matter with $Z_\odot$ should
mix with the matter
that has $Z < Z_{WD}$ and mass fraction of metals in the H-rich envelope of the WD at
the onset of TNR should reduce, while abundance of He should increase, as compared to
the previous outburst. Thus, outburst by outburst, the metals in He+metals layer should
become diluted on a pace set by the mass retained in subsequent outbursts.
These considerations seem to suggest that in mixing models successive H-flashes 
have decreasing strength as a lower C/O abundance is dredged-up so that the corresponding
retention efficiency should increase.
This also implies that a He-buffer massive enough is accumulated, and when it
exceeds a critical value depending on the WD total mass and the effective mass
deposition rate, a He-flash occurs.  

5) For our low-metallicity models, we found that the super-Eddington wind prescription for mass loss by low-mass
WDs results in 100\% retention efficiency $\eta$ (at least for 0.6\,\msun\ WDs),
while it may be as low as 20\% if the RLOF prescription is applied. This
depends, however, on the dimensions of the Roche lobe in a particular binary. For high-mass WDs (1\,\msun), the 
SEW and RLOF prescriptions result in comparable $\eta$ for weak outbursts, but may differ by about $\Delta \log
(\eta) \sim 0.1$ in either direction, depending on the dimensions of the Roche lobe. 
Finding of actual mass-loss algorithm and its efficiency
is one of the most acute problems, which hamper modelling of stellar populations with accreting WDs like CVs or
solving the problem of significance of single-degenerate channel for SNe Ia.  

\section{Acknowledgements}
The authors gratefully acknowldge insightful comments by the anonymous referee which
helped to improve the paper. 
HLC would like to thank Bill Wolf for helpful discussions about  MESA
calculation and Xiangcun Meng for helpful discussion about the influence of
metallicity. We also would like to thank Hans Ritter for helpful discussions on
novae at low metallicity and Oscar Straniero  for discussion on chemical
composition of WDs at different metallicities. We are grateful to the MESA
council for the Mesa instrument papers and website. This work is partially
supported by the National Natural Science Foundation of China (Grants no.
11703081,11521303,11733008), Yunnan Province (No. 2017HC018) and the CAS light
of West China Program, Youth Innovation Promotion Association of Chinese Academy 
of Sciences (Grant no. 2018076). This work was partially supported by Basic Research
Program P-28 of the Presidium of the Russian Academy of Sciences and RFBR
grant No.~19-02-00790. HLC, LRY, TEW gratefully acknowledge  warm
hospitality and support of MPA-Garching. MG acknowledges hospitality of Kazan
Federal University (KFU) and support by the Russian Government Program of
Competitive Growth of KFU. HLC acknowledges the computing time granted by the
Yunnan Observatories and provided on the facilities at the Yunnan Observatories
Supercomputing Platform.



\bibliographystyle{mnras}

\appendix
\label{sec:app} 

\section{Steady burning boundaries for $Z = 0.02$ and $Z = 10^{-4}$}
In table~\ref{tab:snbb_z2m2} and ~\ref{tab:snbb_z1m4}, we present the
steady burning boundaries for different WD mass at $Z = 0.02$ and $Z = 10^{-4}$.

Fitting formulae for the boundaries of steady-burning zones are given by the following formulae.                                                
 These formulae desire the numerical results with the accuracy of better than 0.3\%.

\noindent For $ Z = 0.02$:

\begin{equation}
    {\rm log}(\dot{M}_{\rm lower}) = -10.35 + 8.37 \cdot M_{\rm WD}-6.84 \cdot M_{\rm WD}^{2}+2.07\cdot M_{\rm WD}^{3} 
\end{equation}

\begin{equation}
    {\rm log}(\dot{M}_{\rm upper}) = -9.30 + 6.72 \cdot M_{\rm WD}-5.28 \cdot M_{\rm WD}^{2}+1.50\cdot M_{\rm WD}^{3}  
\end{equation}

\noindent For $Z = 10^{-4}$:
\begin{equation}
    {\rm log}(\dot{M}_{\rm lower}) = -12.21 + 13.32 \cdot M_{\rm WD}-11.90 \cdot M_{\rm WD}^{2}+3.82\cdot M_{\rm WD}^{3}  
\end{equation}

\begin{equation}
    {\rm log}(\dot{M}_{\rm upper}) = -9.81 + 7.25 \cdot M_{\rm WD}-5.17 \cdot M_{\rm WD}^{2}+1.32\cdot M_{\rm WD}^{3}  
\end{equation}

\begin{table*} 
\begin{minipage}[b]{0.98\columnwidth}
\caption{Steady burning boundaries for $Z = 0.02$. The first column is the WD mass; The second and third columns are the lower and upper boundaries, respectively.}
\label{tab:snbb_z2m2}
\begin{center}
\begin{tabular}{ccc}
\hline
    \mwd\; (\msun) &        $\dot{M}_{\rm lower}(M_{\odot}/{\rm yr})$  &     $ \dot{M}_{\rm upper}(M_{\odot}/{\rm yr})$\\
\hline
0.51  &         0.25e-7    &            0.88e-7 \\
0.60  &         0.47e-7    &            1.42e-7 \\
0.70  &         0.73e-7    &            2.14e-7 \\
0.80  &         1.04e-7    &            2.86e-7 \\
0.90  &         1.37e-7    &            3.56e-7 \\
1.00  &         1.73e-7    &            4.26e-7 \\
1.10  &         2.19e-7    &            4.96e-7 \\  
1.20  &         2.65e-7    &            5.60e-7 \\
1.30  &         3.21e-7    &            6.24e-7 \\
\hline
\end{tabular}
\end{center}
\end{minipage}
\hspace{0.04\columnwidth}
\begin{minipage}[b]{0.98\columnwidth}
\caption{Steady burning boundaries for $Z = 10^{-4}$. The first column is the WD mass; The second and third columns are the lower and upper boundaries, respectively.}
\label{tab:snbb_z1m4}
\begin{center}
\begin{tabular}{ccc}
\hline
    \mwd\; (\msun)  &        $\dot{M}_{\rm lower}(M_{\odot}/{\rm yr})$  &     $ \dot{M}_{\rm upper}(M_{\odot}/{\rm yr})$\\
\hline
0.51   &         9.8e-9     &             5.12e-8\\ 
0.60   &         2.13e-8    &             9.36e-8\\  
0.70   &         4.1e-8     &             1.55e-7\\ 
0.80   &         6.0e-8     &             2.18e-7\\ 
0.90   &         8.2e-8     &             3.02e-7\\ 
1.00   &         1.07e-7    &             4.08e-7\\ 
1.10   &         1.39e-7    &             4.54e-7\\ 
1.20   &         1.74e-7    &             5.34e-7\\ 
1.30   &         2.42e-7    &             6.01e-7\\ 
\hline
\end{tabular}
\end{center}
\end{minipage}
\end{table*}

\begin{table*}
    \caption{Characteristics of nova outburst without mixing for solar metallicities. } 
    \label{tab:nova_z2m2}

\begin{tabular}{cccccc}
\hline
$M_{\mathrm{wd}}$ & $\dot{M}_{\mathrm{acc}}$   & Number of & $M_{\rm acc}$ & $M_{\rm ej}$ &  ${\rm log}L_{\rm max}$  \\
    ($M_{\odot}$) & ($\dot{M}_{\odot}/{\rm yr}$) & outburst computed           & ($M_{\odot}$)        & ($M_{\odot}$)    &  ($L_{\odot}$) \\

\hline

0.51 &    1.0e-10&   3    &      3.46e-4           &    1.22e-4                    &   4.06        \\
0.51 &    1.0e-9 &   5    &      2.71e-4           &    1.21e-4                    &   4.06        \\
0.51 &    1.0e-8 &   29   &      1.57e-4           &    1.13e-4                    &   3.90        \\

\hline
0.60 &    1.0e-10&   1    &      2.59e-4           &    2.31e-4                     &   4.65        \\
0.60 &    1.0e-9 &   11   &      2.32e-4           &    2.02e-4                     &   4.66        \\
0.60 &    1.0e-8 &   72   &      1.21e-4           &    1.04e-4                     &   4.09        \\

\hline
0.70 &    1.0e-10&   7    &      2.02e-4           &    1.89e-4                     &   4.54        \\
0.70 &    1.0e-9 &   18   &      1.64e-4           &    1.51e-4                     &   4.66        \\
0.70 &    1.0e-8 &   145  &      8.51e-5           &    7.00e-5                     &   4.23        \\

\hline
0.80 &    1.0e-10&   4    &      1.47e-4           &    1.39e-4                     &   4.55        \\
0.80 &    1.0e-9 &   11   &      1.14e-4           &    1.06e-4                     &   4.56        \\
0.80 &    1.0e-8 &   154  &      5.37e-5           &    4.56e-5                     &   4.36        \\
0.80 &    1.0e-7 &   241  &      1.99e-5           &    0.00                        &   4.31        \\

\hline
0.90 &    1.0e-10&   12   &      1.09e-4           &    1.05e-4                     &   4.70        \\
0.90 &    1.0e-9 &   12   &      7.73e-5           &    7.31e-5                     &   4.57        \\
0.90 &    1.0e-8 &   134  &      3.61e-5           &    3.09e-5                     &   4.43        \\
0.90 &    1.0e-7 &   233  &      1.09e-5           &    0.000                       &   4.41        \\

\hline
1.00 &    1.0e-10&   5    &      6.74e-5           &    6.48e-5                      &   4.84        \\
1.00 &    1.0e-9 &   32   &      5.55e-5           &    5.33e-5                      &   4.51        \\
1.00 &    1.0e-8 &   125  &      2.30e-5           &    2.03e-5                      &   4.51        \\
1.00 &    1.0e-7 &   217  &      6.52e-6           &    2.42e-6                      &   4.49        \\

\hline
1.10 &    1.0e-10&   2    &      4.25e-5           &    4.16e-5                     &   4.71        \\
1.10 &    1.0e-9 &   37   &      3.53e-5           &    3.44e-5                     &   4.58        \\
1.10 &    1.0e-8 &   102  &      1.41e-5           &    1.32e-5                     &   4.58        \\
1.10 &    1.0e-7 &   32   &      3.91e-6           &    2.75e-6                     &   4.56        \\

\hline
1.20 &    1.0e-10&   1    &      2.84e-5           &    2.80e-5                     &   4.66        \\
1.20 &    1.0e-9 &   19   &      1.92e-5           &    1.88e-5                     &   4.64        \\
1.20 &    1.0e-8 &   61   &      6.90e-6           &    6.56e-6                     &   4.64        \\
1.20 &    1.0e-7 &   78   &      2.18e-6           &    1.69e-6                     &   4.62        \\

\hline
1.30 &    1.0e-10&   1    &      1.48e-5           &    1.47e-5                    &   5.20        \\
1.30 &    1.0e-9 &   24   &      7.53e-6           &    7.24e-6                    &   4.75        \\
1.30 &    1.0e-8 &   109  &      2.53e-6           &    2.30e-6                    &   4.72        \\
1.30 &    1.0e-7 &   286  &      8.03e-7           &    6.13e-7                    &   4.70        \\

\hline

\end{tabular}
\end{table*}

\begin{table*}
\caption{Characteristics of nova outburst without mixing for $Z = 0.0001$.  }
\label{tab:nova_z1m4}
\begin{tabular}{cccccc}
\hline
$M_{\mathrm{wd}}$ & $\dot{M}_{\mathrm{acc}}$   & Number of          & $M_{\rm acc}$ & $M_{\rm ej}$  & ${\rm log}L_{\rm max}$  \\
    ($M_{\odot}$)       & ($\dot{M}_{\odot}/{\rm yr}$) & outburst computed  &   ($M_{\odot}$)       & ($M_{\odot}$)            &  ($L_{\odot}$) \\
\hline

0.51 &    1.0e-10 &   93     &      4.28e-4 &   0.000                     & 3.78\\
0.51 &    1.0e-9  &   39     &      4.53e-4 &   0.000                     & 3.76\\

\hline
0.60 &    1.0e-10 &   108    &      3.42e-4 &   0.000                     & 3.98\\
0.60 &    1.0e-9  &   18     &      3.45e-4 &   0.000                     & 3.96\\
0.60 &    1.0e-8  &   177    &      2.16e-4 &   0.000                     & 3.92\\

\hline
0.70 &    1.0e-10 &   13     &      2.59e-4 &   1.27e-4                   & 4.16\\
0.70 &    1.0e-9  &   19     &      2.58e-4 &   7.36e-5                   & 4.14\\
0.70 &    1.0e-8  &   65     &      1.47e-4 &   0.00                      & 4.11\\

\hline
0.80 &    1.0e-10 &   13     &      1.94e-4 &   1.17e-4                   & 4.27\\
0.80 &    1.0e-9  &   17     &      1.94e-4 &   1.11e-4                   & 4.26\\
0.80 &    1.0e-8  &   43     &      1.04e-4 &   0.00                      & 4.24\\

\hline
0.90 &    1.0e-10 &   64     &      1.36e-4 &   8.69e-5                   & 4.38\\
0.90 &    1.0e-9  &   17     &      1.37e-4 &   8.68e-5                   & 4.37\\
0.90 &    1.0e-8  &   140    &      7.24e-5 &   2.63e-5                   & 4.36\\

\hline
1.00 &    1.0e-10 &   64     &      9.47e-5 &   6.98e-5                   & 4.46\\
1.00 &    1.0e-9  &   10     &      9.53e-5 &   7.03e-5                   & 4.46\\
1.00 &    1.0e-8  &   52     &      4.98e-5 &   2.36e-5                   & 4.45\\
1.00 &    1.0e-7  &   79     &      1.62e-5 &   0.00                      & 4.40\\

\hline
1.10 &    1.0e-10 &   9      &      6.36e-5 &   5.24e-5                    & 4.54\\
1.10 &    1.0e-9  &   30     &      6.24e-5 &   5.14e-5                    & 4.54\\
1.10 &    1.0e-8  &   24     &      3.17e-5 &   1.98e-5                    & 4.53\\
1.10 &    1.0e-7  &   41     &      8.61e-6 &   0.00                       & 4.50\\

\hline
1.20 &    1.0e-10 &   39     &      3.89e-5 &   3.45e-5                    & 4.62\\
1.20 &    1.0e-9  &   28     &      3.58e-5 &   3.13e-5                    & 4.61\\
1.20 &    1.0e-8  &   56     &      1.72e-5 &   1.26e-5                    & 4.61\\
1.20 &    1.0e-7  &   123    &      4.12e-6 &   0.00                       & 4.58\\

\hline
1.30 &    1.0e-10 &   2      &      1.93e-5 &   1.85e-5                    & 4.74\\
1.30 &    1.0e-9  &   20     &      1.83e-5 &   1.66e-5                    & 4.70\\
1.30 &    1.0e-8  &   63     &      6.92e-6 &   5.73e-6                    & 4.70\\
1.30 &    1.0e-7  &   84     &      1.43e-6 &   2.18e-7                    & 4.67\\

\hline

\end{tabular}
\end{table*}

\begin{table*}
\caption{Characteristics of nova outburst with mixing for solar metallicities. }
\label{tab:nova_z2m2_mx}
\begin{tabular}{ccccccc}
\hline
$M_{\mathrm{wd}}$ & $\dot{M}_{\mathrm{acc}}$   & number of          & $M_{\rm acc}$ & $M_{\rm ej}$ &  ${\rm log}L_{\rm max}$  \\
    ($M_{\odot}$) & ($\dot{M}_{\odot}/{\rm yr}$) & outburst computed   & ($M_{\odot}$)        & ($M_{\odot}$)     &  ($L_{\odot}$) \\
\hline
0.51 &     1.0e-10 &  1      &      7.10e-5            &   8.88e-5                  &   4.04         \\
0.51 &     1.0e-9  &  46     &      7.88e-5            &   9.67e-5                  &   4.22         \\
0.51 &     1.0e-8  &  71     &      4.28e-5            &   4.63e-5                  &   4.05         \\

\hline
0.60 &     1.0e-10 &  1      &      7.16e-5            &   8.93e-5                  &   4.23         \\
0.60 &     1.0e-9  &  36     &      5.39e-5            &   6.52e-5                  &   4.22         \\
0.60 &     1.0e-8  &  72     &      2.70e-5            &   2.82e-5                  &   4.20         \\

\hline
0.70 &     1.0e-10 &  5      &      5.99e-5            &   7.70e-5                  &   4.36         \\ 
0.70 &     1.0e-9  &  33     &      3.68e-5            &   4.44e-5                  &   4.34         \\
0.70 &     1.0e-8  &  66     &      1.86e-5            &   2.01e-5                  &   4.33         \\
0.70 &     1.0e-7  &  164    &      0.83e-5            &   1.30e-6                  &   4.30         \\

\hline
0.80 &     1.0e-10 &  4      &      4.18e-5            &   5.48e-5                  &   4.43         \\ 
0.80 &     1.0e-9  &  31     &      2.54e-5            &   3.18e-5                  &   4.42         \\
0.80 &     1.0e-8  &  69     &      1.30e-5            &   1.56e-5                  &   4.43         \\
0.80 &     1.0e-7  &  136    &      5.79e-6            &   5.35e-6                  &   4.41         \\

\hline
0.90 &     1.0e-10 &  6      &      2.69e-5            &   3.56e-5                  &   4.76         \\
0.90 &     1.0e-9  &  42     &      1.79e-5            &   2.26e-5                  &   4.52         \\
0.90 &     1.0e-8  &  65     &      8.17e-6            &   9.89e-6                  &   4.50         \\
0.90 &     1.0e-7  &  127    &      3.95e-6            &   4.21e-6                  &   4.49         \\

\hline
1.00 &     1.0e-10 &  8      &      1.96e-5            &   2.56e-5                  &   4.57         \\ 
1.00 &     1.0e-9  &  21     &      9.38e-6            &   1.19e-5                  &   4.57         \\
1.00 &     1.0e-8  &  61     &      5.04e-6            &   6.07e-6                  &   4.58         \\
1.00 &     1.0e-7  &  124    &      2.49e-6            &   2.82e-6                  &   4.56         \\

\hline
1.10 &     1.0e-10 &  7      &      1.67e-5            &   2.22e-5                  &   4.64         \\
1.10 &     1.0e-9  &  41     &      6.39e-6            &   8.32e-6                  &   4.88         \\
1.10 &     1.0e-8  &  63     &      2.64e-6            &   3.27e-6                  &   4.64         \\  
1.10 &     1.0e-7  &  99     &      1.32e-6            &   1.49e-6                  &   4.63         \\

\hline
1.20 &     1.0e-10 &  1      &      2.79e-6            &   3.70e-6                  &   4.87         \\
1.20 &     1.0e-9  &  16     &      2.25e-6            &   2.80e-6                  &   4.72         \\
1.20 &     1.0e-8  &  24     &      1.06e-6            &   1.36e-6                  &   4.69         \\
1.20 &     1.0e-7  &  98     &      6.22e-7            &   6.76e-7                  &   4.69         \\

\hline
1.30 &     1.0e-10 &  2      &      1.02e-6            &   1.24e-6                  &   4.81         \\
1.30 &     1.0e-9  &  11     &      6.02e-7            &   6.69e-7                  &   4.78         \\
1.30 &     1.0e-8  &  30     &      3.68e-7            &   4.11e-7                  &   4.77         \\ 
1.30 &     1.0e-7  &  36     &      1.86e-7            &   1.75e-7                  &   4.75         \\

\hline
\end{tabular}
\end{table*}

\begin{table*}
	\caption{Characteristics of nova outburst with mixing for $Z = 0.0001$. }
\label{tab:nova_z1m4_mx}
\begin{tabular}{cccccccc}
\hline

$M_{\mathrm{wd}}$ & $\dot{M}_{\mathrm{acc}}$   & Number of           & $M_{\rm acc}$ & $M_{\rm ej}$   & ${\rm log}L_{\rm max}$ \\
    ($M_{\odot}$)       & ($\dot{M}_{\odot}/{\rm yr}$) & outburst computed     &   ($M_{\odot}$)       & ($M_{\odot}$)         &  $L_{\odot}$ \\
\hline

0.51 &    1.0e-10  &   3       &     1.10e-4            &   1.21e-4         &   4.10         \\
0.51 &    1.0e-9   &   16      &     7.31e-5            &   7.04e-5         &   4.08         \\
0.51 &    1.0e-8   &   48      &     3.50e-5            &   1.03e-5         &   4.06         \\

\hline
0.60 &    1.0e-10  &   3       &     0.76e-4            &   8.66e-5         &   4.22         \\
0.60 &    1.0e-9   &   31      &     5.35e-5            &   5.59e-5         &   4.21         \\
0.60 &    1.0e-8   &   36      &     2.47e-5            &   1.31e-5         &   4.20         \\

\hline
0.70 &    1.0e-10  &   2       &     4.16e-5            &   4.70e-5         &   4.33         \\ 
0.70 &    1.0e-9   &   32      &     3.48e-5            &   3.79e-5         &   4.34         \\
0.70 &    1.0e-8   &   35      &     1.74e-5            &   1.34e-5         &   4.33         \\
0.70 &    1.0e-7   &   94      &     9.23e-6            &   0.00            &   4.30         \\

\hline
0.80 &    1.0e-10  &   12      &     5.17e-5            &   6.64e-5         &   4.43         \\ 
0.80 &    1.0e-9   &   18      &     3.34e-5            &   4.41e-5         &   4.63         \\
0.80 &    1.0e-8   &   41      &     1.28e-5            &   1.25e-5         &   4.42         \\
0.80 &    1.0e-7   &   121     &     5.42e-6            &   0.00            &   4.40         \\

\hline
0.90 &    1.0e-10  &   17      &     4.52e-5            &   6.01e-5         &   4.51         \\
0.90 &    1.0e-9   &   24      &     1.44e-5            &   1.73e-5         &   4.50         \\
0.90 &    1.0e-8   &   89      &     8.10e-6            &   9.20e-6         &   4.52         \\
0.90 &    1.0e-7   &   71      &     3.74e-6            &   3.16e-6         &   4.49         \\

\hline
1.00 &    1.0e-10  &   6       &     1.69e-5            &   2.25e-5         &   4.73         \\ 
1.00 &    1.0e-9   &   20      &     1.04e-5            &   1.26e-5         &   4.59         \\
1.00 &    1.0e-8   &   41      &     5.03e-6            &   5.80e-6         &   4.58         \\
1.00 &    1.0e-7   &   99      &     2.42e-6            &   2.31e-6         &   4.56         \\

\hline
1.10 &    1.0e-10  &   8       &     1.07e-5            &   1.39e-5         &   4.67         \\
1.10 &    1.0e-9   &   35      &     5.75e-6            &   7.39e-6         &   4.64         \\ 
1.10 &    1.0e-8   &   30      &     2.96e-6            &   3.38e-6         &   4.63         \\
1.10 &    1.0e-7   &   20      &     1.27e-6            &   1.19e-6         &   4.62         \\

\hline
1.20 &    1.0e-10  &   17      &     2.61e-6            &   3.22e-6         &   4.71         \\
1.20 &    1.0e-9   &   88      &     2.06e-6            &   2.49e-6         &   4.70         \\
1.20 &    1.0e-8   &   86      &     1.58e-6            &   1.80e-6         &   4.69         \\
1.20 &    1.0e-7   &   270     &     7.10e-7            &   6.82e-7         &   4.68         \\

\hline
1.30 &    1.0e-10  &   4       &     1.69e-6            &   2.14e-6         &   4.81         \\
1.30 &    1.0e-9   &   31      &     9.53e-7            &   1.12e-6         &   4.78         \\
1.30 &    1.0e-8   &   37      &     5.54e-7            &   6.04e-7         &   4.77         \\
1.30 &    1.0e-7   &   129     &     2.42e-7            &   2.24e-7         &   4.75         \\ 
 
\hline
\end{tabular}
\end{table*}

\bsp
\label{lastpage}

\end{document}